\documentclass[11pt]{article}

\usepackage{RCC}

\title{On Approximating Restricted~Cycle~Covers%
\thanks{Parts of this paper were presented at the 3rd Int.\ Workshop on
Approximation and Online Algorithms
(WAOA 2005)~\cite{Manthey:RestrictedCCWAOA:2006} and the 32nd Int.\ Workshop on
Graph-Theoretic Concepts in Computer Science
(WG 2006)~\cite{Manthey:ImprovedCC:2006}.}}

\author{Bodo Manthey%
\thanks{Work done in part at the Institut f\"ur Theoretische Informatik of the
Universit\"at zu L{\"u}beck supported by German Research Foundation (DFG)
research grant RE 672/3 and at the Department of Computer Science at Yale
University supported by the Postdoc-Program of the German Academic Exchange
Service (DAAD).}}

\date{\small Universit\"at des Saarlandes, Informatik \\
Postfach 151150, 66041 Saarbr\"ucken, Germany \\ 
\texttt{manthey@cs.uni-sb.de}}

\begin{document}
\maketitle

\begin{abstract}
  A cycle cover of a graph is a set of cycles such that every vertex is part of
  exactly one cycle. An $L$-cycle cover is a cycle cover in which the length of
  every cycle is in the set $L$. The weight of a cycle cover of an edge-weighted
  graph is the sum of the weights of its edges.

  We come close to settling the complexity and approximability of computing
  $L$-cycle covers. On the one hand, we show that for almost all $L$, computing
  $L$-cycle covers of maximum weight in directed and undirected graphs is
  \APX-hard. Most of our hardness results hold even if the edge weights are
  restricted to zero and one.

  On the other hand, we show that the problem of computing $L$-cycle covers of
  maximum weight can be approximated within a factor of $2$ for undirected
  graphs and within a factor of $8/3$ in the case of directed graphs. This holds
  for arbitrary sets $L$.
\end{abstract}

%%%%%%%%%%%%%%%%%%%%%%
\section{Introduction}
\label{sec:intro}
%%%%%%%%%%%%%%%%%%%%%%

A cycle cover of a graph is a spanning subgraph that consists solely of cycles
such that every vertex is part of exactly one cycle. Cycle covers play an
important role in the design of approximation algorithms for the traveling
salesman problem~\cite{Blaeser:ATSPZeroOne:2004,BlaeserEA:ATSP:2006,
   BlaeserEA:MetricMaxATSP:2005,BoeckenhauerEA:SharpenedIPL:2000,
   ChandranRam:Parameterized:2007,ChenNagoya:MetricMaxTSP:2007,
   ChenEA:ImprovedMaxTSP:2005,KaplanEA:TSP:2005}, the shortest common
superstring problem~\cite{BlumEA:Superstrings:1994,
   Sweedyk:ApproximationSuperstring:1999}, and vehicle routing
problems~\cite{HassinRubinstein:VehicleRouting:2005}.

In contrast to Hamiltonian cycles, which are special cases of cycle covers,
cycle covers of maximum weight can be computed efficiently. This is exploited in
the aforementioned approximation algorithms, which usually start by computing an
initial cycle cover and then join cycles to obtain a Hamiltonian cycle. This
technique is called \emph{subtour patching}~\cite{GilmoreEA:WellSolved:1985}.

Short cycles in a cycle cover limit the approximation ratios achieved by such
algorithms. In general, the longer the cycles in the initial cover, the better
the approximation ratio. Thus, we are interested in computing cycle covers that
do not contain short cycles. Moreover, there are approximation algorithms that
perform particularly well if the cycle covers computed do not contain cycles of
odd length~\cite{BlaeserEA:ATSP:2006}. Finally, some vehicle routing
problems~\cite{HassinRubinstein:VehicleRouting:2005} require covering vertices
with cycles of bounded length.

Therefore, we consider \emph{restricted cycle covers}, where cycles of certain
lengths are ruled out a priori: For $L \subseteq \nat$, an
\emph{$L$-cycle cover} is a cycle cover in which the length of each cycle is in
$L$. To fathom the possibility of designing approximation algorithms based on
computing cycle covers, we aim to characterize the sets $L$ for which $L$-cycle
covers of maximum weight can be computed, or at least well approximated,
efficiently.

Beyond being a basic tool for approximation algorithms, cycle covers are
interesting in their own right. Matching theory and graph factorization are
important topics in graph theory. The classical matching problem is the problem
of finding one-factors, \ie, spanning subgraphs each vertex of which is incident
to exactly one edge. Cycle covers of undirected graphs are also known as
two-factors because every vertex is incident to exactly two edges. A
considerable amount of research has been done on structural properties of graph
factors and on the complexity of finding graph factors (cf.\ Lov{\'a}sz and
Plummer~\cite{LovaszPlummer:Matching:1986} and
Schrijver~\cite{Schrijver:CombOpt:2003}). In particular, the complexity of
finding restricted two-factors, \ie, $L$-cycle covers in undirected graphs, has
been investigated, and Hell et al.~\cite{HellEA:RestrictedTwoFactors:1988}
showed that finding $L$-cycle covers in undirected graphs is \NP-hard for almost
all $L$. However, almost nothing is known so far about the complexity of finding
directed $L$-cycle covers.

%%%%%%%%%%%%%%%%%%%%%%%%%%
\subsection{Preliminaries}
\label{ssec:prelim}

Let $G=(V,E)$ be a graph with vertex set $V$ and edge set $E$. If $G$ is
undirected, then a \bemph{cycle cover} of $G$ is a subset $C \subseteq E$ of the
edges of $G$ such that all vertices in $V$ are incident to exactly two edges in
$C$. If $G$ is a directed graph, then a cycle cover of $G$ is a subset
$C \subseteq E$ such that all vertices are incident to exactly one incoming and
one outgoing edge in $C$. Thus, the graph $(V,C)$ consists solely of
vertex-disjoint cycles. The length of a cycle is the number of edges it consists
of. Since we do not allow self-loops or multiple edges, the shortest cycles of
undirected and directed graphs are of length three and two, respectively.

We call a cycle of length $\lambda$ a $\lambda$-cycle for short. Cycles of even
or odd length will simply be called even or odd cycles, respectively.

An \bemph{$L$-cycle cover} of an undirected graph is a cycle cover in which the
length of every cycle is in $L \subseteq \unduniv = \{3,4,5,\ldots\}$. An
$L$-cycle cover of a directed graph is analogously defined except that
$L \subseteq \diruniv = \{2,3,4,\ldots\}$. A \bemph{$k$-cycle cover} is a
$\{k, k+1, \ldots\}$-cycle cover. In the following, let
$\oL = \unduniv \setminus L$ in the case of undirected graphs and
$\oL = \diruniv \setminus L$ in the case of directed graphs (whether we consider
undirected or directed cycle covers will be clear from the context).

Given edge weights $w: E \rightarrow \nat$, the \bemph{weight $w(C)$} of a
subset $C \subseteq E$ of the edges of $G$ is $w(C) = \sum_{e \in C} w(e)$. In
particular, this defines the weight of a cycle cover since we view cycle covers
as sets of edges. Let $U \subseteq V$ be any subset of the vertices of $G$. The
\bemph{internal edges of $U$} are all edges of $G$ that have both vertices in
$U$. We denote by \bemph{$w_U(C)$} the sum of the weights of all internal edges
of $U$ that are also contained in $C$. The \bemph{external edges at $U$} are all
edges of $G$ with exactly one vertex in $U$.

For $L \subseteq \unduniv$, \bemph{\ucc L} is the decision problem whether an
undirected graph contains an $L$-cycle cover as spanning subgraph.

\bemph{\maxu L} is the following optimization problem: Given an undirected
complete graph with edge weights zero and one, find an $L$-cycle cover of
maximum weight. We can also consider the graph as being not complete and without
edge weights. Then we try to find an $L$-cycle cover with a minimum number of
``non-edges'' (``non-edges'' correspond to weight zero edges, edges to weight
one edges), \ie, the $L$-cycle cover should contain as many edges as possible.
Thus, \maxu L generalizes \ucc L.

\bemph{\maxug L} is the problem of finding $L$-cycle covers of maximum weight in
graphs with arbitrary non-negative edge weights.

For $k \in \unduniv$, \bemph{\ucc k}, \bemph{\maxu k}, and \bemph{\maxug k} are
defined like \ucc L, \maxu L, and \maxug L except that $k$-cycle covers rather
than $L$-cycle covers are sought.

The problems \bemph{\dcc L}, \bemph{\maxd L}, and \bemph{\maxdg L} as well as
\bemph{\dcc k}, \bemph{\maxd k}, and \bemph{\maxdg k} are defined for directed
graphs like their undirected counterparts except that $L \subseteq \diruniv$ and
$k \in \diruniv$.

An instance of \bemph{\rvc \lambda} is an undirected $\lambda$-regular graph
$H=(X,F)$, \ie, every vertex in $X$ is incident to exactly $\lambda$ edges. A
vertex cover of $H$ is a subset $\tilde X \subseteq X$ such that at least one
vertex of every edge in $F$ is in $\tilde X$. The aim is to find a subset
$\tilde X \subseteq X$ of minimum cardinality. \rvc{\lambda} is \APX-complete
for $\lambda \geq 3$ as follows from results by Alimonti and
Kann~\cite{AlimontiKann:CubicGraphs:2000} as well as Chleb{\'i}k and
Chleb{\'i}kov{\'a}~\cite{ChlebikChlebikova:BoundedOpt:2006}.

An instance of \bemph{\DM \lambda} (exact cover by $\lambda$-sets) is a tuple
$(X, F)$ where $X$ is a finite set and $F$ is a collection of subsets of $X$,
each of cardinality $\lambda$. The question is whether there exists a
sub-collection $\tilde F \subseteq F$ such that for every $x \in X$ there is a
unique $a \in \tilde F$ with $x \in a$. For $\lambda \geq 3$, \DM \lambda\ is
\NP-complete~\cite[SP2]{GareyJohnson:NP:1979}.

Let $\Pi$ be an optimization problem, and let $I$ be its set of instances. For
an instance $X \in I$, let $\opt X$ denote the weight of an optimum solution. We
say that $\Pi$ can be approximated with an approximation ratio of
$\alpha \geq 1$ if there exist a polynomial-time algorithm that, for every
instance $X \in I$, computes a solution $Y$ of $X$ whose weight $w(Y,X)$ is at
most a factor of $\alpha$ away from $\opt X$. This means that
$w(Y,X) \leq \alpha \cdot \opt X$ if $\Pi$ is a minimization problem and
$w(Y,X) \geq \opt X/\alpha$ if $\Pi$ is a maximization
problem~\cite[Definition~3.6]{AusielloEA:ComplApprox:1999}.

%%%%%%%%%%%%%%%%%%%%%%%%%%%%%
\subsection{Previous Results}
\label{ssec:previous}

\maxug{\unduniv}, and thus \ucc{\unduniv} and \maxu{\unduniv}, can be
solved in polynomial time via Tutte's reduction to the classical perfect
matching problem~\cite[Section~10.1]{LovaszPlummer:Matching:1986}. Hartvigsen
presented a polynomial-time algorithm that can be used to decide \ucc 4 in
polynomial time~\cite{Hartvigsen:PhD:1984}. Furthermore, it can be adapted to
solve \maxu 4 as well.

\maxug k admits a simple factor $3/2$ approximation for all $k$: Compute a
maximum weight cycle cover, break the lightest edge of each cycle, and join the
paths thus obtained to a Hamiltonian cycle. Unfortunately, this algorithm cannot
be generalized to work for \maxug L for general $L$. For the problem of
computing $k$-cycle covers of minimum weight in graphs with edge weights one and
two, there exists a factor $7/6$ approximation algorithm for all
$k$~\cite{BlaeserSiebert:CycleCovers:2001}. Hassin and
Rubinstein~\cite{HassinRubinstein:TrianglePacking:2006,
   HassinRubinstein:TrianglePacking:2006Erratum} devised a randomized
approximation algorithm for \maxug{\{3\}} that achieves an approximation ratio
of $83/43 + \epsilon$.

Hell et al.~\cite{HellEA:RestrictedTwoFactors:1988} proved that \ucc L is
\NP-hard for $\oL \not\subseteq \{3,4\}$. For $k\geq 7$, \maxu k and \maxug k
are \APX-complete~\cite{BlaeserManthey:MWCC:2005}. Vornberger showed that
\maxug 5 is \NP-hard~\cite{Vornberger:EasyHard:1980}.

The directed cycle cover problems \dcc \diruniv, \maxd \diruniv, and
\maxdg \diruniv\ can be solved in polynomial time by reduction to the maximum
weight perfect matching problem in bipartite
graphs~\cite[Chapter~12]{AhujaEA:NetworkFlows:1993}. But already \dcc 3 is
\NP-complete~\cite{GareyJohnson:NP:1979}. \maxd k and \maxdg k are \APX-complete
for all $k \geq 3$~\cite{BlaeserManthey:MWCC:2005}.

Similar to the factor $3/2$ approximation algorithm for undirected cycle covers,
\maxdg k has a simple factor $2$ approximation algorithm for all $k$: Compute a
maximum weight cycle cover, break the lightest edge of every cycle, and join the
cycles to obtain a Hamiltonian cycle. Again, this algorithm cannot be
generalized to work for arbitrary $L$. There is a factor $4/3$ approximation
algorithm for \maxdg 3~\cite{BlaeserEA:MetricMaxATSP:2005} and a factor $3/2$
approximation algorithm for \maxd k for
$k \geq 3$~\cite{BlaeserManthey:MWCC:2005}.

The complexity of finding $L$-cycle covers in undirected graphs seems to be well
understood. However, hardly anything is known about the complexity of $L$-cycle
covers in directed graphs and about the approximability of $L$-cycle covers in
both undirected and directed graphs.

%%%%%%%%%%%%%%%%%%%%%%%%
\subsection{Our Results}
\label{ssec:newresults}

We prove that \maxu L is \APX-hard for all $L$ with $\oL \not\subseteq \{3,4\}$
(Section~\ref{subsec:undirecteduniform}) and that \maxug L is \APX-hard if
$\oL \not\subseteq \{3\}$ (Section~\ref{ssec:adaption}). The hardness results
for \maxug L hold even if we allow only the edge weights zero, one, and two.

We show a dichotomy for directed graphs: For all $L$ with $L \neq \{2\}$ and
$L \neq \diruniv$, \dcc L is \NP-hard and \maxd L and \maxdg L are \APX-hard
(Section~\ref{subsec:dirhard}), while all three problems are solvable in
polynomial time if $L = \{2\}$ or $L= \diruniv$.

The hardness results for \maxu L and \maxd L carry over to the problem of
computing $L$-cycle covers of minimum weight in graphs restricted to edge
weights one and two. The hardness results for \maxug L for $\oL = \{3,4\}$ and
$\oL = \{4\}$ carry over to the problem of computing  $L$-cycle covers of
minimum weight where the edge weights are required to fulfill the triangle
inequality.

To show the hardness of directed cycle covers, we show that certain kinds of
graphs, called \emph{$L$-clamps}, exist for non-empty $L \subseteq \diruniv$ if
and only if $L \neq \diruniv$ (Theorem~\ref{thm:dirclampchar}). This
graph-theoretical result might be of independent interest.

Finally, we devise approximation algorithms for \maxug L and \maxdg L that
achieve ratios of $2$ and $8/3$, respectively (Section~\ref{sec:approximation}).
Both algorithms work for all sets~$L$.

%%%%%%%%%%%%%%%%%%%%%%%%%%%%%%%%%%%%%%%%%%%%%%%%%%%%%%%%%%%%%%%%%%
\section{\boldmath The Hardness of Approximating $L$-Cycle Covers}
\label{sec:hardness}
%%%%%%%%%%%%%%%%%%%%%%%%%%%%%%%%%%%%%%%%%%%%%%%%%%%%%%%%%%%%%%%%%%

%%%%%%%%%%%%%%%%%%%%%%%%%%%%%%%
\subsection{Clamps and Gadgets}
\label{subsec:clamps}

To begin the hardness proofs, we introduce \emph{clamps}, which were defined by
Hell et al.~\cite{HellEA:RestrictedTwoFactors:1988}. Clamps are crucial for our
hardness proof.

Let $K=(U,E)$ be an undirected graph, and let $u, v \in U$ be two vertices of
$K$, which we call the \bemph{connectors} of $K$. We denote by $K_{-u}$ and
$K_{-v}$ the graphs obtained from $K$ by deleting $u$ and $v$, respectively, and
their incident edges. $K_{-u-v}$ is obtained from $K$ by deleting both $u$ and
$v$. For $k \in \nat$, $K^k$ is the following graph: Let
$y_1, \ldots, y_k \notin U$ be new vertices, add edges $\{u, y_1\}$,
$\{y_{i}, y_{i+1}\}$ for $1 \leq i \leq k-1$, and $\{y_k, v\}$. For $k = 0$, we
directly connect $u$ to $v$.

Let $L \subseteq \unduniv$. The graph $K$ is called an \bemph{$L$-clamp} if the
following properties hold:
\begin{enumerate}
  \item Both $K_{-u}$ and $K_{-v}$ contain an $L$-cycle cover.
  \item Neither $K$ nor $K_{-u-v}$ nor $K^k$ for any $k \in \nat$ contains an
        $L$-cycle cover.
\end{enumerate}

\begin{figure}[t]
\centering
\subfigure[$L$-clamp.]{\label{fig:finiteclamp}%
\scalebox{0.9}{\includegraphics{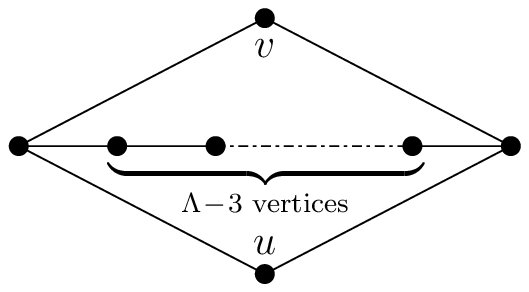}}}
\qquad
\subfigure[$L$-gadget.]{\label{fig:finitegadget}%
\scalebox{0.9}{\includegraphics{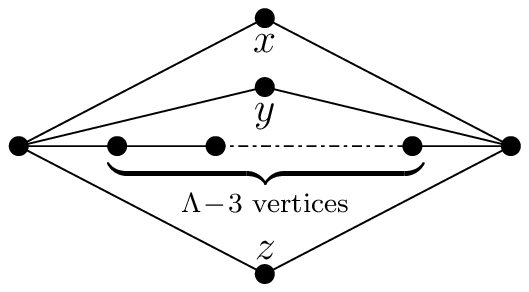}}}
\caption{An $L$-clamp and an $L$-gadget for a set $L$ with $\max(L) = \Lambda$.}
\end{figure}

Figure~\ref{fig:finiteclamp} shows an example of an $L$-clamp for a set $L$ with
$\Lambda = \max(L)$. Hell et al.~\cite{HellEA:RestrictedTwoFactors:1988} proved
the following result which we will exploit for our reduction.

\begin{lemma}[Hell et al.~\cite{HellEA:RestrictedTwoFactors:1988}]
\label{lem:hell}
  Let $L \subseteq \unduniv$ be non-empty. Then there exists an $L$-clamp if and
  only if $\oL \not\subseteq \{3,4\}$.
\end{lemma}

Let $G$ be a graph with vertex set $V$ and $U \subseteq V$. We say that the
vertex set $U$ is an $L$-clamp with connectors $u, v \in U$ in $G$ if the
subgraph of $G$ induced by $U$ is an $L$-clamp and the only external edges of
$U$ are incident to $u$ or $v$.

Let us fix some technical terms. For this purpose, let $C$ be a subset of the
edges of $G$. (In particular, $C$ can be a cycle cover of $G$.) For any
$V' \subseteq V$, we say that \bemph{$V'$ is isolated in $C$} if there is no
edge in $C$ connecting $V'$ to $V \setminus V'$. If $C$ is a cycle cover, then
this means that all cycles of $C$ traverse either only nodes of $V'$ or only
nodes of $V \setminus V'$. We say that the $L$-clamp \bemph{$U$ absorbs $u$ and
expels $v$} if $U \setminus \{v\}$ is isolated in $C$. This means that each
cycle of $C$ traverses either only vertices in $(V \setminus U) \cup \{v\}$ or
only vertices in $U \setminus \{v\}$ (which includes $u$). Analogously,
\bemph{$U$ absorbs $v$ and expels $u$} if $U \setminus \{u\}$ is isolated
in~$C$.

An $L$-clamp implements an exclusive-or of $u$ and $v$: In every $L$-cycle
cover, exactly one of them is absorbed, the other one is expelled. For our
purpose of reducing from \rvc{\lambda}, we need a one-out-of-three behavior. A
graph $K$ is called an \bemph{$L$-gadget with connectors $x,y,z$} if the
following property is fulfilled: Let $G$ be an arbitrary graph that contains $K$
as a subgraph such that only $x$, $y$, and $z$ are incident to external edges.
Then in all $L$-cycle covers $C$ of $G$, exactly two of $K$'s connectors are
expelled while the third one is absorbed. To put it another way: Either
$K_{-x-y}$ or $K_{-x-z}$ or $K_{-y-z}$ is isolated in $C$.

For finite sets $L$, we obtain an $L$-gadget, shown in
Figure~\ref{fig:finitegadget}, by equipping the $L$-clamp of
Figure~\ref{fig:finiteclamp} with an additional connector.

\begin{figure}
\centering
\subfigure[A triple $L$-clamp with connectors $u_1, u_2, u_3$. The connectors of
$L$-clamp $K_i$ are $u_i$ and $v$.]{%
\label{fig:tripleclamp} ~ ~
\scalebox{0.9}{\includegraphics{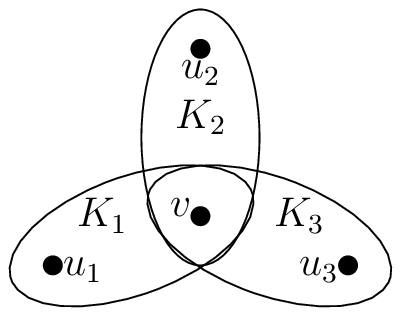}}
 ~ ~ }
\quad
\subfigure[An $L$-gadget with connectors $x,y,z$. The connectors of triple
$L$-clamps $T_i$ are $t_i, u_i, v_i$. For legibility, the triple $L$-clamps are
not shown explicitly but only their connectors.]{%
\label{fig:gadget} ~ \qquad ~ 
\scalebox{0.9}{\includegraphics{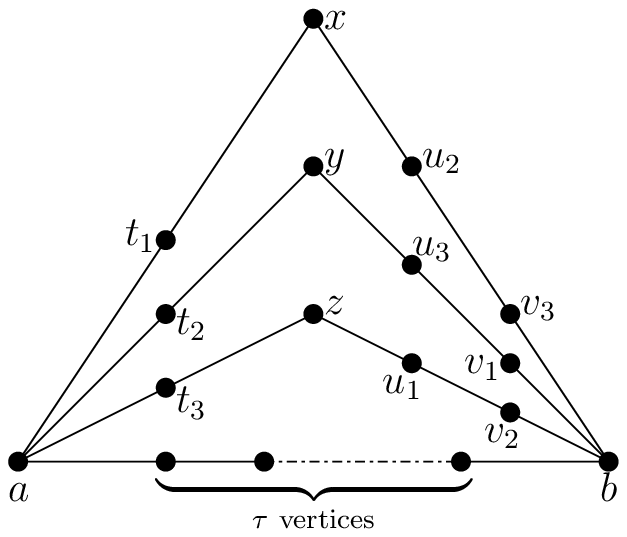}}
~ \qquad ~ }
\caption{A triple $L$-clamp and an $L$-gadget.}
\end{figure}

For infinite sets $L$, we first build an intermediate subgraph. A
\bemph{triple $L$-clamp} is built from three $L$-clamps and has three connectors
$u_1, u_2, u_3$. Figure~\ref{fig:tripleclamp} shows the construction. Triple
$L$-clamps show a two-out-of-three behavior: Only one connector will be
expelled, the other two will be absorbed. More precisely: One of the three
clamps has to absorb $v$. The other two absorb their connectors $u_i$, which are
also connectors of the triple clamp.

Now we are prepared to build $L$-gadgets for infinite sets $L$. These graphs are
built from three triple $L$-clamps $T_1$, $T_2$, and $T_3$, where $T_i$ has
connectors $u_i, v_i, t_i$. Figure~\ref{fig:gadget} shows the $L$-gadget. Since
$L$ is infinite, there exists a $\tau \geq 1$ with $\tau + 6 \in L$. Let us
argue why the $L$-gadget behaves as claimed. For this purpose, let $C$ be an
arbitrary $L$-cycle cover of $G$, where $G$ contains the $L$-gadget as a
subgraph. First, we observe that all $\tau+2$ vertices of the path connecting
$a$ to $b$ must be on the same cycle $c$ in $C$. The only other vertices to
which $a$ is incident are $t_1$, $t_2$, and $t_3$. By symmetry, we assume that
$t_1$ lies also in $c$. Therefore, $T_1$ absorbs $u_1$ and $v_1$. Hence, $v_2$
and $u_3$ are absorbed by $T_2$ and $T_3$, respectively, and $c$ runs through
$x, u_2, v_3$ back to $b$ to form a $(\tau+6)$-cycle. Thus, $x$ is absorbed by
the gadget. $T_2$ expels $u_2$ and absorbs $u_3$, while $T_3$ expels $v_3$ and
absorbs $v_2$. Hence, the gadget expels $y$ and $z$ as claimed. The other two
cases are symmetric.

To conclude this section about clamps, we transfer the notion of $L$-gadgets to
complete graphs with edge weights zero and one and prove some properties. In
Section~\ref{ssec:adaption}, we will generalize the notion of $L$-gadgets to
graphs with arbitrary edge weights.

The transformation to graphs with edge weights zero and one is made in the
obvious way: Let $G$ be an undirected complete graph with vertex set $V$ and
edge weights zero and one. Let $U \subseteq V$. We say that $U$ is an $L$-gadget
with connectors $x,y,z \in U$ if the subgraph of $G$ induced by $U$ restricted
to the edges of weight one is an $L$-gadget with connectors $x, y, z$.

Let $\sigma$ be the number of vertices of an $L$-gadget $U$ with connectors $x$,
$y$, and $z$. Let $C$ be a subset of the edges of $G$ (in particular, $C$ can be
a cycle cover). We call $U$ \bemph{healthy in $C$} if $U$ absorbs $x$, $y$, or
$z$, expels the other two connectors, and $w_U(C) = \sigma -2$. Since the edge
weighted graph is complete, $L$-cycle may traverse $L$-gadgets arbitrarily. The
following lemma shows that we cannot gain weight by not traversing them healthily.

\begin{lemma}
\label{lem:gadgetprop}
  Let $G$ be an undirected graph with vertex set $V$ and edge weights zero and
  one, and let $U \subseteq V$ be an $L$-gadget with connectors $x,y,z$. Let $C$
  be an arbitrary $L$-cycle cover of $G$ and $|U| = \sigma$. Then the following
  properties hold:
  \begin{enumerate}
    \item $w_U(C) \leq \sigma-1$.
    \item If there are $2 \alpha$ external edges at $U$ in $C$, \ie, edges with
          exactly one endpoint in $U$, then $w_U(C) \leq \sigma - \alpha$.
    \item \label{item:legalclamp} Assume that $U$ absorbs exactly one of $x$,
          $y$, or $z$. Then there exists an $L$-cycle cover $\tilde C$ that
          differs from $C$ only in the internal edges of $U$ and has
          $w_U(\tilde C) = \sigma-2$.
    \item \label{item:illegalclamp} Assume that there are two external edges at
          $U$ in $C$ that are incident to two different connectors. Then
          $w_U(C) \leq \sigma-2$.
  \end{enumerate}
\end{lemma}

\begin{proof}
  If $w_U(C) = \sigma$ was true, then $U$ would contain an $L$-cycle cover
  consisting solely of weight one edges since $|U| = \sigma$. This would
  contradict $U$ being an $L$-gadget.

  The second claim follows immediately from $|U| =\sigma$ and the fact that
  every vertex is incident to exactly two edges in a cycle cover.

  Assume without loss of generality that $U$ absorbs $x$ and expels $y$ and $z$.
  Since $U$ is an $L$-gadget, $U \setminus \{y,z\}$ contains an $L$-cycle cover
  consisting of $\sigma-1$ weight one edges, which proves the third claim.

  The fourth claim remains to be proved. If there are more than two external
  edges at $U$ in $C$, we have at least four external edges and thus
  $w_U(C) \leq \sigma -2$. So assume that there are exactly two external edges
  at $U$ in $C$ incident to, say, $x$ and $y$. We have $\sigma-1$ internal edges
  of $U$ in $C$. If all of them had weight one, this would contradict the
  property that in an unweighted $L$-gadget always $U \setminus \{x,y\}$,
  $U \setminus \{x,z\}$, or $U \setminus \{y,z\}$ is isolated.
\end{proof}

%%%%%%%%%%%%%%%%%%%%%%%%%%%%%%%%%%%%%%%%%%%%%%%%
\subsection{The Reduction for Undirected Graphs}
\label{subsec:undirecteduniform}

The notion of L-reductions was introduced by Papadimitriou and
Yannakakis~\cite{PapadimitriouYannakakis:Optimization:1991} (cf.\ Ausiello et
al.~\cite[Definition~8.4]{AusielloEA:ComplApprox:1999}). L-reductions can be
used to show the \APX-hardness of optimization problems. We present an
L-reduction from \rvc \lambda\ to show the inapproximability of \maxu L for
$\oL \not\subseteq \{3,4\}$. The inapproximability of \maxug L for
$\oL \not\subseteq \{3\}$ and \maxd L for $L \neq \{2\}$ and $L \neq \diruniv$
will be shown in subsequent sections.

Let $L \subseteq \unduniv$ be non-empty with $\oL \not\subseteq \{3,4\}$. Thus,
$L$-gadgets exist and we fix one as in the previous section. Let
$\lambda = \min(L)$. (This choice is arbitrary. We could choose any number in
$L$.) We will reduce \rvc \lambda\ to \maxu L. $\rvc \lambda$ is \APX-complete
since $\lambda \geq 3$.

Let $H=(X,F)$ be an instance of \rvc \lambda\ with $|X| = n$ vertices and
$|F| = m = \lambda n/2$ edges. Our instance $G$ for \maxu L consists of
$\lambda$ subgraphs $G_1, \ldots, G_\lambda$, each containing $\sigma m$
vertices, where $\sigma$ is the number of vertices of the $L$-gadget. We start
by describing $G_1$. Then we state the differences between $G_1$ and
$G_2, \ldots, G_\lambda$ and say to which external edges of
$G_1, \ldots, G_\lambda$ weight one is assigned.

Let $a = \{x,y\} \in F$ be any edge of $H$. We construct an $L$-gadget $F_a$ for
$a$ that has connectors $x_a^1$, $y_a^1$ and $z_a^1$. We call $F_a$ an
\bemph{edge gadget}.

Now let $x \in X$ be any vertex of $H$ and let $a_1, \ldots, a_\lambda \in F$ be
the $\lambda$ edges that are incident to $x$. We connect the vertices
$x_{a_1}^1, \ldots, x_{a_\lambda}^1$ to form a path by assigning weight one to
the edges $\{x_{a_\eta}^1, x_{a_{\eta+1}}^1\}$ for
$\eta \in \{1, \ldots, \lambda-1\}$. Together with edge
$\{x_{a_\lambda}^1, x_{a_1}^1\}$, these edges form a cycle of length
$\lambda \in L$, but note that $w(\{x_{a_\lambda}^1, x_{a_1}^1\}) = 0$. These
$\lambda$ edges are called the \bemph{junctions of $x$}. The
\bemph{junctions at $F_a$} for some $a = \{x, y\} \in F$ are the junctions of
$x$ and $y$ that are incident to $F_a$. Overall, the graph $G_1$ consists of
$\sigma m$ vertices since every edge gadget consists of $\sigma$ vertices.

The graphs $G_2, \ldots, G_\lambda$ are almost exact copies of $G_1$. The graph
$G_\xi$ ($\xi \in \{2,\ldots, \lambda\}$) consists of $L$-gadgets with
connectors $x_a^\xi$, $y_a^\xi$, and $z_a^\xi$ for each edge
$a = \{x,y\} \in F$, just as above. The edge weights are also identical with the
single exception that the edge $\{x_{a_\lambda}^\xi, x_{a_1}^\xi\}$ also has
weight one. Note that we use the term ``edge gadget'' only for the subgraphs
$F_a$ of $G_1$ defined above although almost the same subgraphs occur in
$G_2, \ldots, G_\lambda$ as well. Similarly, the term ``junction'' refers only
to edges in $G_1$.

\begin{figure}[t]
\centering 
\scalebox{0.9}{\includegraphics{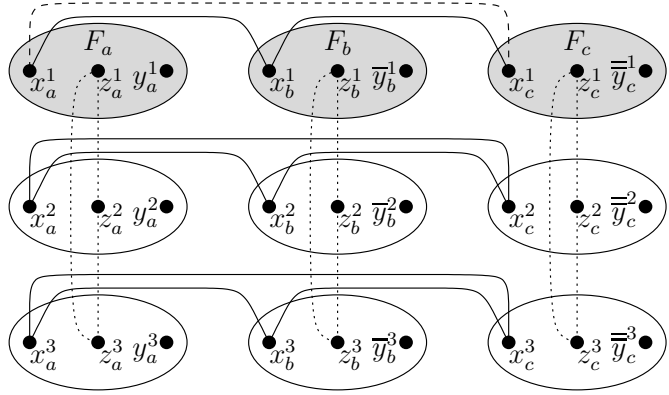}}
\caption{The construction for $x \in X$ incident to $a = \{x,y\},
b = \{x,\overline y\}, c=\{x,\overline{\overline{y}}\} \in F$ for $\lambda = 3$.
$F_a$, $F_b$, and $F_c$ are grey. The three ellipses in the second and third row
build $G_2$ and $G_3$, respectively. The cycles connecting the $z$-vertices are
dotted. The junctions of $x$ and their copies are solid, except for
$\{x_c^1, x_a^1\}$, which has weight zero and is dashed.}
\label{fig:example}
\end{figure}

Finally, we describe how to connect $G_1, \ldots, G_\lambda$ with each other.
For every edge $a \in F$, there are $\lambda$ vertices
$z_a^1, \ldots, z_a^\lambda$. These are connected to form a cycle consisting
solely of weight one edges, \ie, we assign weight one to all edges
$\{z_a^\xi, z_a^{\xi+1}\}$ for $\xi \in \{1, \ldots, \lambda-1\}$ and to
$\{z_a^\lambda, z_a^1\}$. Figure~\ref{fig:example} shows an example of the whole
construction from the viewpoint of a single vertex.

Edges with both vertices in the same gadget are called \bemph{internal edges}.
Besides junctions and internal edges, the third kind of edges are the
\bemph{$z$-edges} of $F_a$ for $a \in F$, which are the two edges
$\{z_a^1, z_a^2\}$ and $\{z_a^1, z_a^\lambda\}$. The fourth kind of edges are
\bemph{illegal edges}, which are edges that are not junctions but connect any
two vertices of two different gadgets. The $z$-edges, however, are not illegal.
Edges within $G_2, \ldots, G_\lambda$ as well as edges connecting $G_\xi$ to
$G_{\xi'}$ for $\xi, \xi' \geq 2$ have no special name.

We define the following terms for arbitrary subsets $C$ of the edges of the
graph $G$ thus constructed, which includes the case of $C$ being a cycle cover.
Let $a = \{x,y\} \in F$ be an arbitrary edge of $H$. We say that
\bemph{$C$ legally connects $F_a$} if the following properties are fulfilled:
\begin{itemize}
  \item $C$ contains either two or four of the junctions at $F_a$ and no illegal
        edges incident to~$F_a$.
  \item If $C$ contains exactly two junctions at $F_a$, then these belong to the
        same vertex and the two $z$-edges at $F_a$ are contained in $C$.
  \item If $C$ contains four junctions at $F_a$, then $C$ does not contain the
        $z$-edges at $F_a$.
\end{itemize}
We call $C$ \bemph{legal} if $C$ legally connects all gadgets. If $\tilde C$ is
a legal $L$-cycle cover, then for all $x\in X$ either all junctions of $x$ or no
junction of $x$ is in $\tilde C$. From a legal $L$-cycle cover $\tilde C$, we
obtain the subset $\tilde X = \{x\mid \text{the junctions of $x$ are in
$\tilde C$}\} \subseteq X$. Since at least two junctions at $F_a$ are in
$\tilde C$ for every $a \in F$, the set $\tilde X$ is a vertex cover of $H$.

The idea behind the reduction is as follows: Consider an edge
$a = \{x,y\} \in F$. We interpret $x_a^1$ being expelled to mean that $x$ is in
the vertex cover. (In this case, the junctions of $x$ are in the cycle cover.)
Analogously, $y$ is in the vertex cover if $y_a^1$ is expelled. The vertex
$z_a^1$ is only absorbed if both $x$ and $y$ are in the vertex cover. If only
one of $x$ and $y$ is in the vertex cover, $z_a^1$ forms a $\lambda$-cycle
together with $z_a^2,\ldots, z_a^\lambda$.

We only considered $G_1$ when defining the terms ``legally connected'' and
``legal.'' This is because in $G_1$, we lose weight one for putting $x$ into the
vertex cover since the junction $\{x_{a_\lambda}^1, x_{a_1}^1\}$ weighs zero.
The other $\lambda -1$ copies of the construction are only needed because
$z_a^1$ must be part of some cycle if $z_a^1$ is not absorbed.

\begin{lemma}
\label{lem:vctocc}
  Let $\tilde X$ be a vertex cover of size $\tilde n$ of $H$. Then $G$ contains
  an $L$-cycle cover $\tilde C$ with
  $w(\tilde C) = \sigma \lambda m - \tilde n$.
\end{lemma}

\begin{proof}
  We start by describing $\tilde C$ in $G_1$. For every vertex $x \in \tilde X$,
  the cycle consisting of all $\lambda$ junctions is in $\tilde C$. Let
  $a = \{x,y\} \in F$ be any edge. Then either $x$ or $y$ or both are in
  $\tilde X$. If only $x$ is in $\tilde X$, we let $F_a$ absorb $y_a^1$ while
  $z_a^1$ is expelled. If only $y$ is in $\tilde X$, we let $F_a$ absorb $x_a^1$
  while $z_a^1$ is again expelled. If both $x$ and $y$ are in $\tilde X$, then
  we let $x_a^1$ and $y_a^1$ be expelled while $z_a^1$ is absorbed.

  We perform the same construction as for $G_1$ for all copies
  $G_2, \ldots, G_\lambda$. If $z_a^1$ is expelled, then
  $z_a^2, \ldots, z_a^\lambda$ are expelled as well. We let them form a
  $\lambda$-cycle in $\tilde C$.

  Clearly, $\tilde C$ is legal. Furthermore, $\tilde C$ is an $L$-cycle cover:
  Every cycle either has a length of $\lambda \in L$ or lies totally inside a
  single $L$-gadget. All $L$-gadgets are healthy in $\tilde C$, thus $\tilde C$
  is an $L$-cycle cover.

  All edges of $\tilde C$ within $G_2, \ldots, G_\lambda$ have weight one. The
  only edges that connect different copies $G_\xi$ and $G_{\xi'}$ are edges
  $\{z_a^{\xi}, z_a^{\xi+1}\}$ and $\{z_a^\lambda, x_a^1\}$, which have weight
  one as well. Almost all edges used in $G_1$ also have weight one; the only
  exception is one junction of weight zero for each $x \in \tilde X$. Since
  $|\tilde X| = \tilde n$, there are $\tilde n$ edges of weight zero in
  $\tilde C$. The graph $G$ contains $\sigma \lambda m$ vertices, thus
  $\tilde C$ contains $\sigma \lambda m$ edges, $\sigma \lambda m - \tilde n$ of
  which have weight one.
\end{proof}

Let $C$ be an $L$-cycle cover of $G$ and let $a \in F$. We define $W_{F_a}(C)$
as the sum of the weights of all internal edges of $F_a$ plus half the number of
$z$-edges in $C$ at $F_a$. Analogously, $W_{G_\xi}(C)$ is the number of weight
one edges with both vertices in $G_\xi$ plus half the number of weight one edges
with exactly one vertex in $G_\xi$.

\begin{lemma}
\label{lem:weightcharac}
  Let $C$ be an $L$-cycle cover and let $j$ be the number of weight one
  junctions in $C$. Then
  $w(C) = j + \sum_{a \in F} W_{F_a}(C) + \sum_{\xi = 2}^\lambda W_{G_\xi}(C)$.
\end{lemma}

\begin{proof}
  Every edge with both vertices in the same $G_\xi$ is counted once. The only
  edges of weight one between different $G_\xi$ are the edges
  $\{z_a^\xi, z_a^{\xi+1}\}$ and $\{z_a^\lambda, z_a^1\}$. These are counted
  with one half in both $W_{G_\xi}(C)$ and $W_{G_{\xi+1}}(C)$ for
  $2 \leq \xi \leq \lambda-1$ or one half in both $W_{G_\xi}(C)$ and
  $W_{F_a}(C)$ for $\xi \in \{2,\lambda\}$.
\end{proof}

In a legal $L$-cycle cover $\tilde C$ as described in Lemma~\ref{lem:vctocc}, we
have $W_{G_\xi}(\tilde C) = \sigma m$ for all $\xi \in \{2, \ldots, \lambda\}$
since every vertex in $G_\xi$ is only incident to edges of weight one in
$\tilde C$ by construction. Now we show that it is always best to traverse the
gadgets legally and to keep the gadgets healthy.

\begin{lemma}
\label{lem:generallegal2}
  Given an arbitrary $L$-cycle cover $C$, we can compute a legal $L$-cycle cover
  $\tilde C$ with $w(\tilde C) \geq w(C)$ in polynomial time.
\end{lemma}

\begin{proof}
  We proceed as follows to obtain $\tilde C$:
  \begin{enumerate}
    \item Let $C'$ be $C$ with all illegal edges removed.
    \item For all $x \in X$ in arbitrary order: If at least one junction of $x$
          is in $C$, then put all junctions of $x$ into $C'$.
    \item \label{step:lonely} For all $a = \{x,y\} \in F$ in arbitrary order: If
          neither the junctions of $x$ nor the junctions of $y$ are in $C'$,
          choose arbitrarily one vertex of $a$, say $x$, and add all junctions
          of $x$ to $C'$.
    \item Rearrange $C'$ within $G_1$ such that all clamps are healthy in $C'$.
    \item Rearrange $C'$ such that all $G_2, \ldots, G_\lambda$ are traversed
          exactly like $G_1$.
    \item For all $a \in F$: If $z_a^1, \ldots, z_a^\xi$ are not absorbed, let
          them form a $\lambda$-cycle. Call the result~$\tilde C$.
  \end{enumerate}
  The running-time of the algorithm is polynomial. Moreover, $\tilde C$ is a
  legal $L$-cycle cover by construction. What remains is to prove
  $w(\tilde C) \geq w(C)$.

  Let $w(C) = j + \sum_{a \in F} W_{F_a}(C) +
  \sum_{\xi = 2}^{\lambda} W_{G_\xi}(C)$ be the weight of $C$ according to
  Lemma~\ref{lem:weightcharac}, \ie, $C$ contains $j$ junctions of weight one.
  Analogously, let $w(\tilde C) = \tilde{\jmath} +
  \sum_{a \in F} W_{F_a}(\tilde C) + \sum_{\xi = 2}^{\lambda}
  W_{G_\xi}(\tilde C)$, \ie, $\tilde \jmath$ is the number of junctions of
  weight one in $\tilde C$.

  All illegal edges have weight zero, and we do not remove any junctions. We
  have $W_{G_\xi}(\tilde C) = \sigma m$ for all $\xi$, which is maximal. Thus,
  no weight is lost in this way. What remains is to consider the internal edges
  of the gadgets and the $z$-edges.

  Let $a = \{x, y\}$ be an arbitrary edge of $H$. If
  $W_{F_a}(C) \leq W_{F_a}(\tilde C)$, then nothing has to be shown. Those
  gadgets $F_a$ with $W_{F_a}(C) > W_{F_a}(\tilde C)$ remain to be considered.
  We have $W_{F_a}(\tilde C) \geq \sigma -2$ and $W_{F_a}(C) \leq \sigma -1$
  according to Lemma~\ref{lem:gadgetprop}. Thus, $W_{F_a}(C) =  \sigma -1$ and
  $W_{F_a}(\tilde C) =  \sigma -2 = W_{F_a}(C) -1$ for all $a \in F$ with
  $W_{F_a}(C) > W_{F_a}(\tilde C)$. What remains to be proved is that for all
  such gadgets, there is a junction of weight one in $\tilde C$ that is not in
  $C$ and can thus compensate for the loss of weight one in $F_a$. This means
  that we have to show that $\tilde \jmath$ is at least $j$ plus the number of
  edges $a$ with $W_{F_a}(C)>W_{F_a}(\tilde C)$.

  If $W_{F_a}(C) = \sigma -1$, then according to
  Lemma~\ref{lem:gadgetprop}(\ref{item:illegalclamp}), the junctions at $F_a$ in
  $C$ (if there are any) belong to the same vertex. Since
  $W_{F_a}(\tilde C) =  \sigma -2$, all four junctions at $F_a$ are in
  $\tilde C$. Thus, while executing the above algorithm, there is a moment at
  which at least one of, say, $y$'s junctions at $F_a$ is in $C'$, and the
  junctions of $x$ are added in the next step. We say that a vertex
  \bemph{$x$ compensates $F_a$} if
  \begin{enumerate}
    \item $\tilde C$ contains $x$'s junctions,
    \item no junction of $x$ at $F_a$ is in $C$, and 
    \item at the moment at which $x$'s junctions are added, $C'$ already
          contains at least one junction of $y$ at $F_a$.
  \end{enumerate}
  Thus, every gadget $F_a$ with $W_{F_a}(\tilde C) < W_{F_a}(C)$ is compensated
  by some vertex $x \in a$.

  It remains to be shown that the number of gadgets that are compensated by some
  vertex is at most equal to the number of weight one junctions added to $C'$.
  Let $\eta \in \{0, \ldots, \lambda\}$ be the number of junctions of $x$ in
  $C$. If $\eta = \lambda$, then $x$ does not compensate any gadget. If
  $\eta = 0$, \ie, $C$ does not contain any of $x$'s junctions, then the
  junctions of $x$ are added during Step~\ref{step:lonely} of the algorithm
  because there is some edge $a \in F$ with $x \in a$ such that there is no
  junction at all in $C'$ at $F_a$ before adding $x$'s junctions. Thus, $x$ does
  not compensate $F_a$. At most $\lambda -1$ gadgets are compensated by $x$, and
  $\lambda-1$ junctions of $x$ have weight one. The case that remains is
  $\eta \in \{1, \ldots, \lambda-1\}$. Then $\lambda - \eta$ junctions of $x$
  are added and at least $\lambda - \eta -1$ of them have weight one. On the
  other hand, there are at least $\eta + 1$ gadgets $F_a$ such that at least one
  junction of $x$ at $F_a$ is already in $C$: Every junction is at two gadgets,
  and thus $\eta$ junctions are at $\eta +1$ or more gadgets. Thus, at most
  $\lambda - \eta -1$ gadgets are compensated by $x$.
\end{proof}

Finally, we prove the following counterpart to Lemma~\ref{lem:vctocc}.

\begin{lemma}
\label{lem:getset2}
  Let $\tilde C$ be the $L$-cycle cover constructed as described in the proof of
  Lemma~\ref{lem:generallegal2} and let
  $\tilde X = \{x \mid \text{$x$'s junctions are in $\tilde C$}\}$ be the subset
  of $X$ obtained from $\tilde C$. Choose $\tilde n$ such that
  $w(\tilde C) = \sigma \lambda m - \tilde n$. Then $|\tilde X| = \tilde n$.
\end{lemma}

\begin{proof}
  The proof is similar to the proof of Lemma~\ref{lem:vctocc}. We set the weight
  of all junctions to one. With respect to the modified edge weights, the weight
  of $\tilde C$ is $\sigma \lambda m$. Thus, $\tilde n$ is the number of weight
  zero junctions in $\tilde C$, which is just $|\tilde X|$.
\end{proof}

Now we are prepared to prove the main theorem of this section.

\begin{theorem}
\label{thm:Lapx}
  For all $L \subseteq \unduniv$ with $\oL \not\subseteq \{3,4\}$, \maxu L is
  \APX-hard.
\end{theorem}

\begin{proof}
  We show that the reduction presented is an L-reduction. Then the result
  follows from the \APX-hardness of \rvc{\lambda}. Let $\opt{H}$ be the size of
  a minimum vertex cover of $H$ and $\opt{G}$ be the weight of a maximum weight
  $L$-cycle cover of $G$. From Lemmas~\ref{lem:vctocc}, \ref{lem:generallegal2},
  and~\ref{lem:getset2}, we obtain that
  $\opt G =  \sigma \lambda m - \opt H \leq  \sigma \lambda m$. Since $H$ is
  $\lambda$-regular, we have $\opt{H} \geq n/2$. Thus, 
  \[
  \opt{G} \leq \sigma \lambda m
          =    \sigma \lambda \cdot (\lambda n/2)
          \leq (\sigma \lambda^2) \cdot \opt{H} .
  \]

  Let $C$ be an arbitrary $L$-cycle cover of $G$, $\tilde C$ be a legal
  $L$-cycle cover obtained from $C$ as in Lemma~\ref{lem:generallegal2}, and
  $\tilde X \subseteq X$ obtained from $\tilde C$. Then
  \[
       \bigl| |\tilde X| - \opt H \bigr|
  =    \bigl| w(\tilde C) - \opt G \bigr|
  \leq \bigl| w(C) - \opt G\bigr| ,
  \]
  which completes the proof.
\end{proof}

%%%%%%%%%%%%%%%%%%%%%%%%%%%%%%%%%%%%%%%%%%%%%%%%%%%%%%%%%%%%
\subsection{\boldmath Adaption of the Reduction to \maxug L}
\label{ssec:adaption}

To prove the \APX-hardness of \maxug L for $\oL \not\subseteq \{3\}$, all we
have to do is to deal with $\oL = \{4\}$ and $\oL = \{3,4\}$. For all other sets
$L$, the inapproximability follows from Theorem~\ref{thm:Lapx}. We will adapt
the reduction presented in the previous section.

\begin{figure}[t]
\centering \subfigure[The clamp.]{%
\label{fig:weightedclamp} \quad
\scalebox{0.9}{\includegraphics{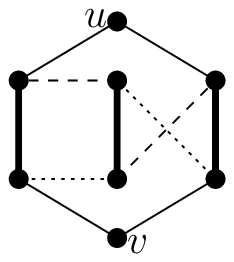}}
\quad}
 ~
\subfigure[Absorbing $v$.]{%
\label{fig:w1} \quad
\scalebox{0.9}{\includegraphics{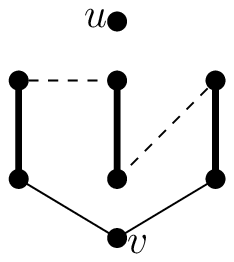}}
\quad}
 ~
\subfigure[Absorbing $u$.]{%
\label{fig:w2} \quad
\scalebox{0.9}{\includegraphics{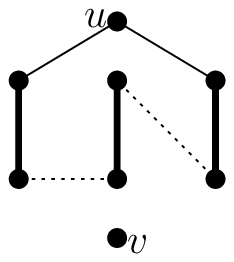}}
\quad}
 ~
\subfigure[Illegal traversal of $K_x$.]{%
\label{fig:weightillegal}
\quad \quad ~
\scalebox{0.9}{\includegraphics{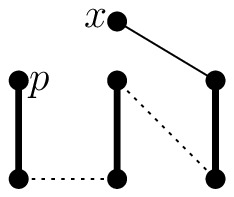}
\quad \quad ~}}
\caption{A weighted $L$-clamp for $\{4\} \subseteq \oL \subseteq \{3,4\}$ and
how to traverse it. Bold edges have weight two; solid, dashed, and dotted edges
have weight one.}
\end{figure}

To do this, we have to find an edge weighted analog of an $L$-clamp. We do not
explicitly define the properties a weighted $L$-clamp has to fulfill. Instead,
we just call the graph shown in Figure~\ref{fig:weightedclamp} a
\bemph{weighted $L$-clamp} for $\oL =\{3,4\}$ and $\oL =\{4\}$.

The basic idea is that all three edges of weight two of the weighted clamp have
to be traversed in a cycle cover. Since $4$-cycles are forbidden, we have to
take either the two dotted edges or the two dashed edges. Otherwise, we would
have to take an edge of weight zero. Furthermore, if we take the dashed edges,
we have to absorb $v$ and to expel $u$, and if we take the dotted edges, we have
to absorb $u$ and to expel $v$ (Figures~\ref{fig:w1} and~\ref{fig:w2}). Again,
we would have to take edges of weight zero otherwise.

\begin{figure}[t]
\centering\subfigure[The weighted $L$-gadget.]{%
\label{fig:weightedgadget}\scalebox{0.9}{\includegraphics{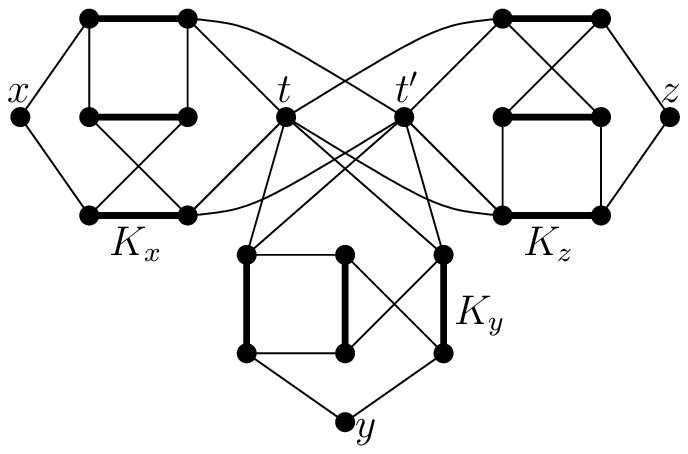}}}
\qquad 
\subfigure[How to absorb $x$.]{%
\label{fig:weightedabsorb}\scalebox{0.9}{\includegraphics{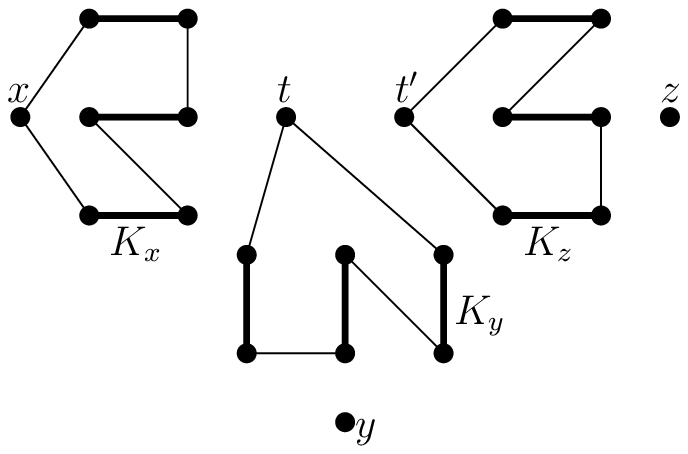}}}
\caption{A weighted $L$-gadget and how to use it.}
\end{figure}

Using three weighted $L$-clamps $K_x, K_y, K_z$, we build an $L$-gadget as shown
in Figure~\ref{fig:weightedgadget}. Note that both $t$ and $t'$ can serve as a
connector for each of the clamps. This weighted $L$-gadget has essentially the
same properties as the $L$-gadgets of Section~\ref{subsec:clamps}, which were
stated as Lemma~\ref{lem:gadgetprop}. The difference is that $\sigma = 32$ is no
longer the number of vertices, but the number of vertices plus the number of
edges of weight two.

\begin{lemma}
  Let $G$ be an undirected graph with vertex set $V$ and edge weights zero and
  one, and let $U \subseteq V$ be a weighted $L$-gadget with connectors $x,y,z$
  in $G$. Let $C$ be an arbitrary $L$-cycle cover of $G$. Then the following
  properties hold:
  \begin{enumerate}
    \item $w_U(C) \leq 31$.
    \item If there are $2 \alpha$ external edges at $U$ in $C$, then
          $w_U(C) \leq 32-\alpha$.
    \item If $U$ absorbs $x$, then there exists an $L$-cycle cover $\tilde C$
          that differs from $C$ only in the internal edges of $U$ and has
          $w_U(\tilde C) = 30$. The same holds if $U$ absorbs $y$ or $z$.
    \item Assume that there are two external edges at $U$ in $C$ that are
          incident to two different connectors. Then $w_U(C) \leq 30$.
  \end{enumerate}
\end{lemma}

\begin{proof}
  The only way to achieve $w_U(C) > 31$ is $w_U(C) =32$, which requires that we
  have $23$ internal edges including all nine edges of weight two. Since
  $4$-cycles are forbidden, such an $L$-cycle cover does not exist.

  If we have $2\alpha$ external edges, then we have $23-\alpha$ internal edges.
  At most nine of them are of weight two.

  If $U$ absorbs $x$, then we can achieve a weight of 30 by letting $K_y$ and
  $K_z$ absorb $t_1$ and $t_2$, respectively (Figure~\ref{fig:weightedabsorb}).
  (We can also connect $K_y$ and $K_z$ via $t$ and $t'$ to obtain a $14$-cycle.
  The weight would be the same.) In the same way, we can achieve weight 30 if
  $U$ absorbs $y$ or $z$.

  The fourth claim remains to be proved. We have $w_U(C) \leq 31$ and $22$
  internal edges. If $w_U(C) > 30$, then $w_U(C) = 31$, and $C$ contains all
  nine edges of weight two and no internal edge of weight zero of $U$. By
  symmetry, it suffices to consider the case that $x$ is incident to one
  external edge. Figure~\ref{fig:weightillegal} shows which edges are mandatory
  in order to keep all three edges of weight two. Since the cycle that contains
  $x$ must be continued at $p$, vertex $p$ is incident to an edge of weight zero
  in $C$, which proves the claim.
\end{proof}

Given these properties, we can plug the $L$-gadget into the reduction described
in the previous section to obtain the \APX-hardness of $\maxug L$ for
$\oL = \{4\}$ and $\oL = \{3,4\}$. Together with Theorem~\ref{thm:Lapx}, we
obtain the following result.

\begin{theorem}
\label{thm:maxug}
  \maxug{L} is \APX-hard for all $L$ with $\oL \not\subseteq \{3\}$ even if the
  edge weights are restricted to be zero, one, or two.
\end{theorem}

%%%%%%%%%%%%%%%%%%%%%%%%%%%%%%%%%%%%%%
\subsection{Clamps in Directed Graphs}
\label{subsec:dirclamps}

The aim of this section is to prove a counterpart to Lemma~\ref{lem:hell} (for
the existence of $L$-clamps) for directed graphs. Let $K=(V,E)$ be a directed
graph and $u,v \in V$. Again, $K_{-u}$, $K_{-v}$, and $K_{-u-v}$ denote the
graphs obtained by deleting $u$, $v$, and both $u$ and $v$, respectively. For
$k \in \nat$, $K^k_u$ denotes the following graph: Let
$y_1, \ldots, y_k \notin V$ be new vertices and add edges
$(u, y_1), (y_1, y_2), \ldots, (y_k, v)$. For $k=0$, we add the edge $(u,v)$.
The graph $K^{k}_v$ is similarly defined, except that we now start at $v$, \ie,
we add the edges $(v, y_1), (y_1, y_2), \ldots, (y_k, u)$. $K^{0}_v$ is $K$ with
the additional edge $(v,u)$.

Now we can define clamps for directed graphs: Let $L\subseteq \diruniv$. A
directed graph $K=(V,E)$ with $u,v \in V$ is a \bemph{directed $L$-clamp} with
connectors $u$ and $v$ if the following properties hold:
\begin{itemize}
  \item Both $K_{-u}$ and $K_{-v}$ contain an $L$-cycle cover.
  \item Neither $K$ nor $K_{-u-v}$ nor $K^k_u$ nor $K^k_v$ for any $k \in \nat$
        contains an $L$-cycle cover.
\end{itemize}

Let us now prove that directed $L$-clamps exist for almost all $L$.

\begin{theorem}
\label{thm:dirclampchar}
  Let $L \subseteq \diruniv$ be non-empty. Then there exists a directed
  $L$-clamp if and only if $L \neq \diruniv$.
\end{theorem}

\begin{figure}[t]
\centering 
\subfigure[A $\{2\}$-clamp.]{%
\label{fig:dir1clamp}\scalebox{0.9}{\includegraphics{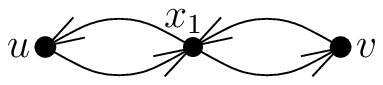}}}
\quad 
\subfigure[An $L$-clamp for finite sets $L$ with $\max(L) = \Lambda \geq 3$.]{%
\label{fig:dirfinclamp}\scalebox{0.9}{\includegraphics{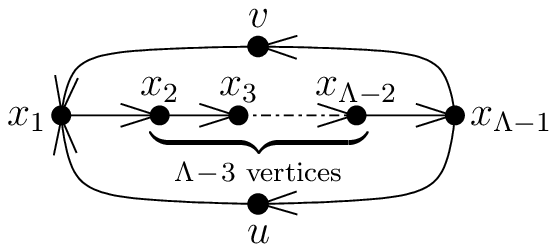}}}
\quad
\subfigure[A $\overline{\{2\}}$-clamp.]{%
\label{fig:dir2clamp}\scalebox{0.9}{\includegraphics{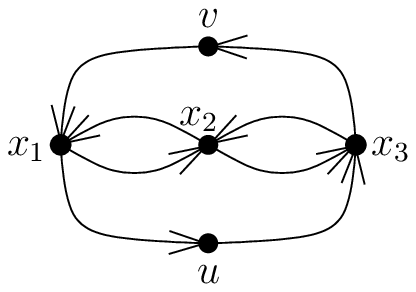}}}
\\
\subfigure[An $L$-clamp for $\Lambda \not\in L$ and $\Lambda+2 \in L$ with
$\Lambda \geq 3$.]{%
\label{fig:dirinfclamp2}\scalebox{0.9}{\includegraphics{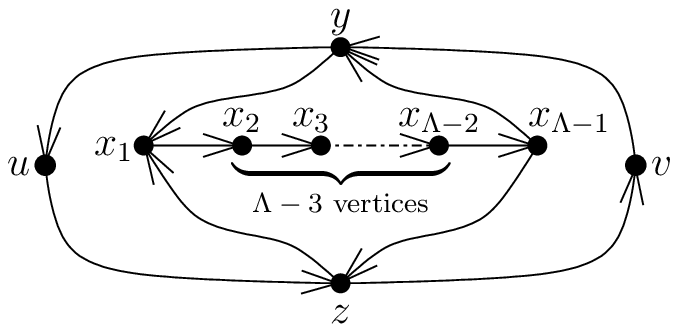}}}
\quad
\subfigure[An $L$-clamp for $\Lambda, \Lambda+2 \not\in L$ and
$\Lambda + 1 \in L$ with $\Lambda \geq 4$.]{%
\label{fig:dirinfclamp1}\scalebox{0.9}{\includegraphics{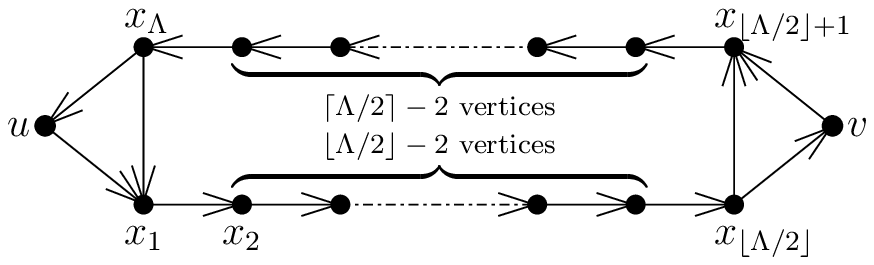}}}
\caption{Directed $L$-clamps. The connectors are $u$ and $v$, the internal
vertices are $x_1, x_2, \ldots$ and $y,z$.}
\label{fig:directedclamps}
\end{figure}

\begin{proof}
  We first prove that directed $L$-clamps exist for all non-empty sets
  $L \subseteq \diruniv$ with $L \neq \diruniv$. We start by considering finite
  $L$. If $L$ is finite, $\max(L) = \Lambda$ exists. For $L = \{2\}$, the graph
  shown in Figure~\ref{fig:dir1clamp} is a directed $L$-clamp: Either $u$ or $v$
  forms a $2$-cycle with $x_1$, and there are no other possibilities. Otherwise,
  we have $\Lambda \geq 3$. Figure~\ref{fig:dirfinclamp} shows a directed
  $L$-clamp for this case, which is a directed variant of the undirected clamp
  shown in Figure~\ref{fig:finiteclamp}.

  Now we consider finite $\oL$. Figure~\ref{fig:dir2clamp} shows an $L$-clamp
  for $\oL = \{2\}$: $x_1$, $x_2$, and $x_3$ must be on the same path since
  length two is forbidden. This cycle must include $u$ or $v$ but cannot include
  both of them

  Otherwise, $\max(\oL) = \Lambda \geq 3$ and $\Lambda +2 \in L$ and the graph
  shown in Figure~\ref{fig:dirinfclamp2} is an $L$-clamp: The vertices
  $x_1, \ldots, x_{\Lambda-1}$ must all be on the same cycle. Thus, either
  $(y, x_1)$ or $(z, x_1)$ is in the cycle cover. By symmetry, it suffices to
  consider the first case. Since $\Lambda \notin L$, the edge
  $(x_{\Lambda-1}, y)$ cannot be in the cycle cover. Thus, $(v,y)$ and
  $(x_{\Lambda-1},z)$ and hence $(z,v)$ are in the cycle cover.

  The case that remains to be considered is that both $L$ and $\oL$ are
  infinite. We distinguish two sub-cases. Either there exists a
  $\Lambda \geq 4$ with $\Lambda, \Lambda +2 \notin L$ and $\Lambda +1\in L$. In
  this case, the graph shown in Figure~\ref{fig:dirinfclamp1} is an $L$-clamp:
  $x_1, \ldots, x_\Lambda$ must be on the same cycle. Since the lengths
  $\Lambda$ and $\Lambda +2$ are not allowed, either $v$ or $u$ is expelled and
  the other vertex is absorbed.

  If no $\Lambda$ exists with $\Lambda, \Lambda +2 \notin L$ and
  $\Lambda+1 \in L$ (but $L$ and $\oL$ are infinite), then there exists a
  $\Lambda \geq 3$ with $\Lambda \notin L$ and $\Lambda +2 \in L$ and we can use
  the graph already used for finite $\oL$ (Figure~\ref{fig:dirinfclamp2}) as a
  directed $L$-clamp.

  Lemma~\ref{lem:alsocc} below shows that $\diruniv$-clamps do not exist, which
  completes the proof.
\end{proof}

\begin{lemma}
\label{lem:alsocc}
  Let $G=(V,E)$ be a directed graph and let $u, v\in V$. If $G_{-u}$ and
  $G_{-v}$ both contain a cycle cover, then
  \begin{itemize}
    \item both $G$ and $G_{-u-v}$ contain cycle covers or
    \item all $G^k_u$ and $G^k_v$ for $k \in \nat$ contain cycle covers.
  \end{itemize}
\end{lemma}

\begin{proof}
  Let $E_{-u}$ and $E_{-v}$ be the sets of edges of the cycle covers of $G_{-u}$
  and $G_{-v}$, respectively. We construct two sequences of edges
  $P = (e_1, e_2, \ldots)$ and $P' =(e'_1, e'_2, \ldots)$. These sequences can
  be viewed as augmenting paths and we use them to construct cycle covers of
  $G_{-u-v}$ and $G$ or $G_u^k$ and $G_v^k$. The sequence $P$ is given uniquely
  by traversing edges of $E_{-v}$ forwards and edges of $E_{-u}$ backwards:
  \begin{itemize}
    \item $e_1=(u, x_1)$ is the unique outgoing edge of $u = x_0$ in $E_{-v}$.
    \item If $e_i = (x_{i-1}, x_i) \in E_{-v}$, \ie, if $i$ is odd, then 
          $e_{i+1} =(x_{i+1}, x_i) \in E_{-u}$ is the unique incoming edge of
          $x_i$ in $E_{-u}$.
    \item If $e_i = (x_{i}, x_{i-1}) \in E_{-u}$, \ie, if $i$ is even, then
          $e_{i+1}=(x_i, x_{i+1}) \in E_{-v}$ is the unique outgoing edge of
          $x_{i}$ in $E_{-v}$.
    \item If in any of the above steps no extension of $P$ is possible, then
          stop.
  \end{itemize}

  Let $P = (e_1, \ldots, e_\ell)$. We observe two properties of the sequence
  $P$.

\begin{lemma}
\label{lem:propertiesaug}
  \begin{enumerate}
    \item No edge appears more than once in $P$.
    \item If $\ell$ is odd, \ie, $e_\ell \in E_{-v}$, then
          $e_\ell = (x_{\ell-1}, u)$. If $\ell$ is even, \ie,
          $e_\ell \in E_{-u}$, then $e_\ell = (v, x_{\ell-1})$.
  \end{enumerate}
\end{lemma}

\begin{proof}
  Assume the contrary of the first claim and let $e_i = e_j$ ($i \neq j$) be an
  edge that appears at least twice in $P$ such that $i$ is minimal. If $i=1$,
  then $e_j =(u, x_1) \in E_{-v}$. This would imply
  $e_{j-1} = (u, x_{j-2}) \in E_{-u}$, a contradiction. If $i>1$, then assume
  $e_i=(x_{i-1}, x_i) \in E_{-v}$ without loss of generality. Since exactly one
  edge leaves $x_{i-1}$ in $E_{-u}$, the edge $e_{i-1} = e_{j-1}$ is uniquely
  determined, which contradicts the assumption that $i$ be minimal.

  Let us now prove the second claim. Without loss of generality, we assume that
  the last edge $e_\ell$ belongs to $E_{-v}$. Let
  $e_\ell = (x_{\ell-1}, x_\ell)$. The path $P$ cannot be extended, which
  implies that there does not exist an edge $(x_{\ell+1}, x_\ell) \in E_{-u}$.
  Since $E_{-u}$ is a cycle cover of $G_{-u}$, this implies $x_\ell = u$ and
  completes the proof.
\end{proof}

  We build the sequence $P'$ analogously, except that we start with the edge
  $e'_1 = (x'_1, v) \in E_{-u}$. Again, we traverse edges of $E_{-v}$ forwards
  and edges of $E_{-u}$ backwards. Let $P' = (e'_1, \ldots, e'_{\ell'})$.

  No edge appears in both $P$ and $P'$ as can be proved similarly to the first
  claim of Lemma~\ref{lem:propertiesaug}. Moreover, either $P$ ends at $u$ and
  $P'$ ends at $v$ or vice versa: We have $e_\ell = (x_{\ell-1}, u)$ if and only
  if $e'_{\ell'} = (v, x_{\ell'-1})$, and we have $e_\ell = (v, x_{\ell-1})$ if
  and only if $e'_{\ell'} = (x_{\ell'-1}, u)$. Let $P_{-u} \subseteq E_{-u}$
  denote the set of edges of $E_{-u}$ that are part of $P$. The sets $P_{-v}$,
  $P'_{-u}$, $P'_{-v}$ are defined similarly.

\begin{figure}[t]
\centering
\subfigure[A graph $G$.]{%
\label{gcc:subfig:agraph}%
\parbox{0.46\textwidth}{\centering \includegraphics{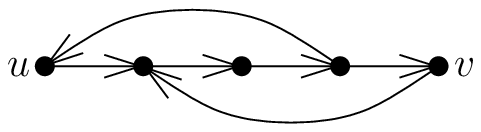}}}
\qquad 
\subfigure[Cycle covers of $G_{-v}$ (dashed and solid) and $G_{-u}$ (dotted and
solid).]{%
\label{gcc:subfig:acc}%
\parbox{0.46\textwidth}{\centering \includegraphics{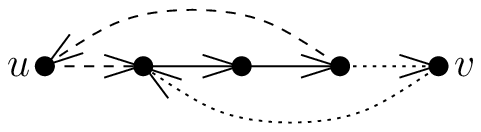}}}
\\
\subfigure[$P$ (top) and $P'$ (bottom). Dashed and dotted edges belong to the
cycle covers of $G_{-v}$ and $G_{-u}$, respectively.]{%
\label{gcc:subfig:ap}%
\parbox{0.46\textwidth}{\centering \includegraphics{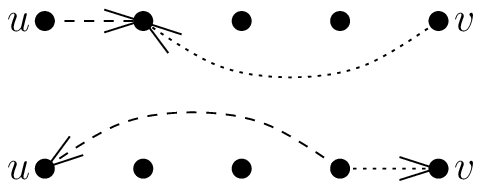}}}
\qquad 
\subfigure[Cycle covers of $G_v^0$ (top) and $G_u^0$ (bottom).]{%
\label{gcc:subfig:anew}%
\parbox{0.46\textwidth}{\centering \includegraphics{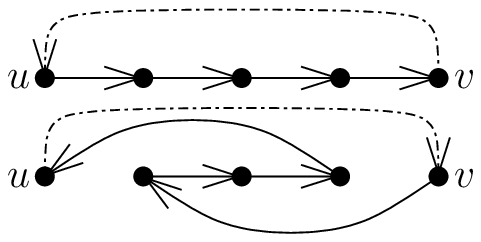}}}
\caption{Constructing cycle covers of $G_v^0$ and $G_u^0$ from the sequences $P$
and $P'$.}
\label{gcc:fig:aug}
\end{figure}

\begin{figure}[t]
\centering
\subfigure[Another graph $G$.]{%
\label{gcc:subfig:bgraph}%
\parbox{0.46\textwidth}{\centering \includegraphics{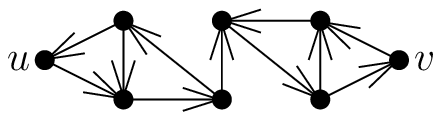}}}
\qquad
\subfigure[Cycle covers of $G_{-v}$ (dashed and solid) and $G_{-u}$ (dotted and
solid).]{%
\label{gcc:subfig:bcc}%
\parbox{0.46\textwidth}{\centering \includegraphics{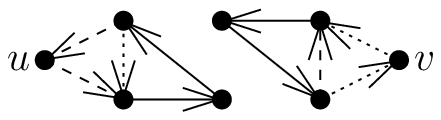}}}
\\
\subfigure[$P$ (top) and $P'$ (bottom).]{%
\label{gcc:subfig:bp}%
\parbox{0.46\textwidth}{\centering \includegraphics{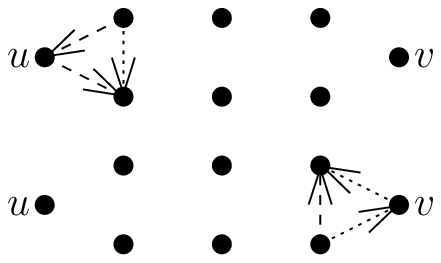}}}
\qquad
\subfigure[Cycle covers of $G$ (top) and $G_{-u-v}$ (bottom).]{%
\label{gcc:subfig:bnew}%
\parbox{0.46\textwidth}{\centering \includegraphics{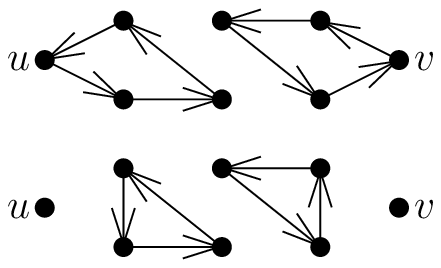}}}
\caption{Constructing cycle covers of $G$ and $G_{-u-v}$ from the sequences $P$
and $P'$.}
\label{gcc:fig:bug}
\end{figure}

  Two examples are shown in Figures~\ref{gcc:fig:aug} and~\ref{gcc:fig:bug}:
  Figures~\ref{gcc:subfig:agraph} and~\ref{gcc:subfig:acc} show a graph with its
  cycle covers, while Figure~\ref{gcc:subfig:ap} depicts $P$ and $P'$, the
  former starting at $u$ and ending at $v$ and the latter starting at $v$ and
  ending at $u$. Figures~\ref{gcc:subfig:bgraph}, \ref{gcc:subfig:bcc},
  and~\ref{gcc:subfig:bp} show another example graph, this time $P$ starts and
  ends at $u$ and $P'$ starts and ends at $v$.

  Our aim is now to construct cycle covers of $G$ and $G_{-u-v}$ or of $G_u^k$
  and $G_v^k$. We distinguish two cases. Let us start with the case that $P$
  starts at $u$ and ends at $v$ and, consequently, $P'$ starts at $v$ and ends
  at $u$. Then
  \[
  E_u^0 = (E_{-v} \setminus P_{-v}) \cup P_{-u} \cup \{(u,v)\}
  \]
  is a cycle cover of $G_u^0$. To prove this, we have to show
  $\indeg_{E_u^0}(x) = \outdeg_{E_u^0}(x) = 1$ for all $x \in V$:
  \begin{itemize}
    \item We removed the outgoing edge of $u$ in $E_{-v}$, which is in $P_{-v}$.
          The incoming edge of $u$ in $E_{-v}$ is left. $P_{-u}$ does not
          contain any edge incident to $u$ and $(u,v)$ is an outgoing edge of
          $u$. Thus, $\indeg_{E_u^0}(u) = \outdeg_{E_u^0}(u) = 1$.
    \item There is no edge incident to $v$ in $E_{-v}$. $P_{-u}$ contains an
          outgoing edge of $v$ and $(u,v)$ is an incoming edge of $v$. Thus,
          $\indeg_{E_u^0}(v) = \outdeg_{E_u^0}(v) = 1$.
    \item For all $x \in V \setminus \{u,v\}$, either both $P_{-v}$ and $P_{-u}$
          contain an incoming edge of $x$ or none of them does. Analogously,
          either both $P_{-v}$ and $P_{-u}$ contain an outgoing edge of $x$ or
          none of them does. Thus, replacing $P_{-v}$ by $P_{-u}$ changes
          neither $\indeg(x)$ nor $\outdeg(x)$.
  \end{itemize}
  By replacing the edge $(u,v)$ by a path $(u, y_1), \ldots, (y_k, v)$, we
  obtain a cycle cover of $G_u^k$ for all $k \in \nat$. A cycle cover of $G_v^0$
  is obtained similarly:
  \[
  E_v^0 = (E_{-u} \setminus P_{-u}) \cup P_{-v} \cup \{(v,u)\}  .
  \]
  As above, we get cycle covers of $G_v^k$ by replacing $(v,u)$ by a path
  $(v, y_1), \ldots, (y_k, u)$. Figure~\ref{gcc:subfig:anew} shows an example
  how the new cycle covers are obtained.

  The second case is that $P$ starts and ends at $u$ and $P'$ starts and ends at
  $v$. Then
  \[
  (E_{-v} \setminus P_{-u}) \cup P_{-v} \: \text{ and} \:
  (E_{-u} \setminus P'_{-v}) \cup P'_{-u}
  \]
  are cycle covers of $G$ and 
  \[
  (E_{-v} \setminus P_{-v}) \cup P_{-u} \: \text{ and} \:
  (E_{-u} \setminus P'_{-u}) \cup P'_{-v}
  \]
  are cycle covers of $G_{-u-v}$. The proof is similar to the first case.
  Figure~\ref{gcc:subfig:bnew} shows an example.
\end{proof}

%%%%%%%%%%%%%%%%%%%%%%%%%%%%%%%%%%%%%%%%%%%%%%%%%%%%%%%%%
\subsection{\boldmath Intractability for Directed Graphs}
\label{subsec:dirhard}

From the hardness results in the previous sections and the work by Hell et
al.~\cite{HellEA:RestrictedTwoFactors:1988}, we obtain the \NP-hardness and
\APX-hardness of \dcc L and \maxd L, respectively, for all $L$ with $2 \notin L$
and $\oL \not\subseteq \{2,3,4\}$: We use the same reduction as for undirected
cycle covers and replace every undirected edge $\{u,v\}$ by a pair of directed
edges $(u,v)$ and $(v,u)$. However, this does not work if $2 \in L$ and also
leaves open the cases when $\oL \subsetneq \{2,3,4\}$. \dcc{\diruniv},
\maxd{\diruniv}, and \maxdg{\diruniv} can be solved in polynomial time, but the
case $L = \{2\}$ is also easy: Replace two opposite edges $(u,v)$ and $(v,u)$ by
an edge $\{u,v\}$ of weight $w(u,v) + w(v,u)$ and compute a matching of maximum
weight on the undirected graph thus obtained.

We will settle the complexity of the directed cycle cover problems by showing
that $L = \{2\}$ and $L = \diruniv$ are the only tractable cases. For all other
$L$, \dcc L is  \NP-hard and \maxd L and \maxdg L are \APX-hard. Let us start by
proving the \APX-hardness.

\begin{theorem}
\label{thm:apxdir}
  Let $L \subseteq \diruniv$ be a non-empty set. If
  $L \notin \{\{2\}, \diruniv\}$, then \maxd L is \APX-hard.
\end{theorem}

\begin{proof}
  We adapt the proof presented in Section~\ref{subsec:undirecteduniform}. Since
  $L \neq \{2\}$, there exists a $\lambda \in L$ with $\lambda \geq 3$. Thus,
  $\rvc \lambda$ is \APX-complete. All we need is such a $\lambda$ and a
  directed $L$-clamp. Then we can reduce \rvc \lambda\ to \maxd L.

  We use the $L$-clamps to build $L$-gadgets, which again should have the
  property that they absorb one of their connectors and expel the other two. In
  case of $L$ being finite, the graph shown in Figure~\ref{fig:dirfingadget} is
  a directed $L$-gadget. In case of infinite $L$, we can build directed triple
  $L$-clamps exactly as for undirected graphs. Using these, we can build
  directed $L$-gadgets, which are simply directed variants of their undirected
  counterparts (Figure~\ref{fig:dirinfgadget}).

\begin{figure}[t]
\centering
\subfigure[$L$-gadget for finite $L$.]{%
\label{fig:dirfingadget}%
\scalebox{0.9}{\includegraphics{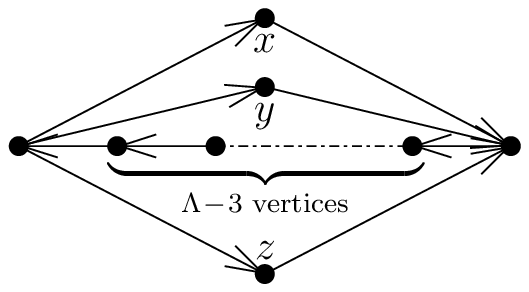}}}
\qquad
\subfigure[$L$-gadget for infinite $L$ with $\tau + 6 \in L$. The triple clamps
are represented by their connectors $t_i, u_i, v_i$.]{%
\label{fig:dirinfgadget}%
\qquad
\scalebox{0.9}{\includegraphics{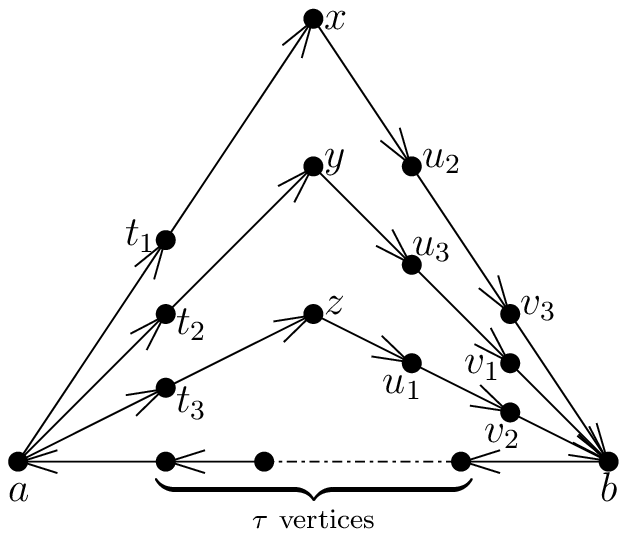}}
\qquad}
\caption{Directed $L$-gadgets with connectors $x,y,z$.}
\end{figure}

  The edge gadgets build the graph $G_1$: Let $x \in X$ be a vertex of $H$ and
  $a_1, \ldots, a_\lambda \in F$ be the edges incident to $x$ in $H$ (in
  arbitrary order). Then we assign weight one to the edges
  $(x_{a_\xi}^1, x_{a_{\xi+1}}^1)$ for all $\xi \in \{1, \ldots, \lambda-1\}$.
  The edge $(x_{a_\lambda}^1, x_{a_1}^1)$ has weight zero. These $\lambda$ edges
  are called the junctions of $x$.

  Again, $G_2, \ldots, G_\lambda$ are exact copies of $G_1$ except that weight
  one is assigned also to $(x_{a_\lambda}^\xi, x_{a_1}^\xi)$ for all
  $\xi \in \{2,3,\ldots, \lambda\}$.

  Again, we let the $z$-vertices form $\lambda$-cycles: For all edges $a \in F$,
  we assign weight one to $(z_a^\xi, z_a^{\xi+1})$ for
  $\xi \in \{1,2, \ldots, \lambda-1\}$ and to $(z_a^\lambda, z_a^1)$.

  Weight zero is assigned to all edges that are not mentioned.

  The remainder of the proof goes along the same lines as the \APX-hardness
  proof for undirected $L$-cycle covers.
\end{proof}

Note that the \NP-hardness of \dcc L for $L \notin \{\{2\}, \diruniv\}$ does not
follow directly from the \APX-hardness of \maxd L: A famous counterexample is
\prob{2SAT}, for which it is \APX-hard to maximize the number of simultaneously
satisfied clauses~\cite{PapadimitriouYannakakis:Optimization:1991}, although
testing whether a 2CNF formula is satisfiable takes only linear time.

\begin{theorem}
\label{thm:npdir}
  Let $L \subseteq \diruniv$ be a non-empty set. If
  $L \notin \{\{2\}, \diruniv\}$, then \dcc L is \NP-hard.
\end{theorem}

\begin{proof}
  All we need is an $L$-clamp and some $\lambda \in L$ with $\lambda \geq 3$. We
  present a reduction from $\DM \lambda$ (which is \NP-complete since
  $\lambda \geq 3$) that is similar to the reduction of Hell et
  al.~\cite{HellEA:RestrictedTwoFactors:1988} used to prove the \NP-hardness of
  \ucc L for $\oL \not\subseteq \{3,4\}$.

  Let $(X, F)$ be an instance of \DM \lambda. Note that we will construct a
  directed graph $G$ as an instance of $\dcc L$, \ie, $G$ is neither complete
  nor edge-weighted. For each $x \in X$, we have a vertex in $G$ that we again
  call $x$. For $a = \{x_1, \ldots, x_\lambda\} \in F$, we construct a
  $\lambda$-cycle consisting of the vertices $a_1, \ldots, a_\lambda$. Then we
  add $\lambda$ $L$-clamps $K_a^{x_\eta}$ with $a_\eta$ and $x_\eta$ as
  connectors for all $\eta \in \{1, \ldots, \lambda\}$. See
  Figure~\ref{fig:exampledirnp} for an example.

\begin{figure}[t]
\centering \scalebox{0.9}{\includegraphics{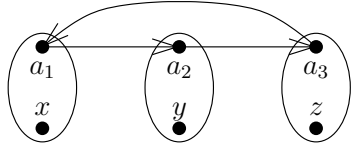}}
\caption{The construction for the \NP-hardness of \dcc L\ from the viewpoint of
$a = \{x,y,z\} \in F$. Each ellipse represents an $L$-clamp.}
\label{fig:exampledirnp}
\end{figure}

  What remains to be shown is that $G$ contains an $L$-cycle cover if and only
  if $F$ is a ``yes'' instance of $\DM \lambda$. Assume first that there exists
  a subset $\tilde F \subseteq F$ such that $\bigcup_{a \in \tilde F} a = X$ and
  every element $x \in X$ is contained in exactly one set of $\tilde F$. We
  construct an $L$-cycle cover of $G$ in which all clamps are healthy: Let
  $a = \{x_1, \ldots, x_\lambda\} \in F$. If $a \in \tilde F$, then let
  $K_a^{x_\eta}$ expel $a_\eta$ and absorb $x_\eta$ for all
  $\eta \in \{1, \ldots, \lambda\}$, and let $a_1, a_2, \ldots, a_\lambda$ form
  a $\lambda$-cycle. If $a \notin \tilde F$, let $K_a^{x_\eta}$ expel $x_\eta$
  and absorb $a_\eta$ for all $\eta \in \{1, \ldots, \lambda\}$. All connectors
  are absorbed by exactly one clamp or are covered by a $\lambda$-cycle since
  $\tilde F$ is an exact cover.

  Now we prove the reverse direction. Let $C$ be an $L$-cycle cover of $G$. Then
  every clamp of $G$ is healthy in $C$, \ie, it absorbs one of its connectors
  and expels the other one. Let $a = \{x_1, \ldots, x_\lambda\} \in F$ and
  assume that $K_a^{x_\eta}$ expels $a_\eta$. Since $a_\eta$ must be part of a
  cycle in $C$, $(a_{\eta-1}, a_\eta)$ and $(a_\eta, a_{\eta+1})$ must be in
  $C$. We obtain that either all $a_1, \ldots, a_\lambda$ are absorbed by
  $K_a^{x_1}, \ldots, K_a^{x_\lambda}$ or that all are expelled by
  $K_a^{x_1}, \ldots, K_a^{x_\lambda}$. Now consider any $x \in X$ and let
  $a_1, a_2, \ldots, a_\ell \in F$ be all the sets that contain $x$. All clamps
  $K_{a_1}^x, \ldots, K_{a_\ell}^x$ are healthy, $C$ is an $L$-cycle cover of
  $G$, and $x$ is not incident to any further edges. Hence, there must be a
  unique $a_i$ such that $K_{a_i}^x$ absorbs $x$. Thus,
  \[
  \tilde F = \bigl\{ a = \{x_1, \ldots, x_\lambda\} \in F \mid
  \text{$K_a^{x_\eta}$ absorbs $x_\eta$ for all
        $\eta \in \{1, \ldots, \lambda\}$}  \bigr\}
  \]
  is an exact cover of $(X,F)$.
\end{proof}

If the language $\{1^\lambda\mid \lambda \in L\}$ is in \NP, then \dcc L is also
in \NP\ and therefore \NP-complete if $L \notin \{\{2\}, \diruniv\}$: We can
nondeterministically guess a cycle cover and then check if $\lambda \in L$ for
every cycle length $\lambda$ occurring in that cover. Conversely, if
$\{1^\lambda\mid \lambda \in L\}$ is not in $\NP$, then \dcc L is not in \NP\
either since there is a reduction of $\{1^\lambda\mid \lambda \in L\}$ to
\dcc L: On input $x = 1^\lambda$, construct a graph $G$ on $\lambda$ vertices
that consists solely of a Hamiltonian cycle. Then $x \in L$ if and only if $G$
contains an $L$-cycle cover.

%%%%%%%%%%%%%%%%%%%%%%%%%%%%%%%%%%
\section{Approximation Algorithms}
\label{sec:approximation}
%%%%%%%%%%%%%%%%%%%%%%%%%%%%%%%%%%

The goal of this section is to devise approximation algorithms for \maxug L and
\maxdg L that work for arbitrary $L$. The catch is that we have an uncountable
number of problems $\maxug L$ and $\maxdg L$ and for most $L$ it is impossible
to decide whether some cycle length is in $L$ or not.

Assume, for instance, that we have an algorithm that solves \maxug L for some
set $L$ that is not recursively enumerable. We enumerate all instances of
\maxug L and run the algorithm on these instances. This yields an enumeration of
a subset of $L$. Since $L$ is not recursively enumerable, there exist
$\lambda \in L$ such that the algorithm never outputs $\lambda$-cycles. Now
consider a graph with $\lambda$ vertices where all edges have weight zero except
for a Hamiltonian cycle of weight one edges. Then the Hamiltonian cycle is the
unique optimum solution, but our algorithm does not output the $\lambda$-cycle,
contradicting the assumption it solves $\maxug L$.

One possibility to circumvent this problem would be to restrict ourselves to
sets $L$ such that $\{1^\lambda \mid \lambda \in L\}$ is in \DP. Another
possibility to cope with this problem is to include the permitted cycle lengths
in the input. However, while such restrictions are necessary for finding optimum
solutions, it turns out that they are unnecessary for designing approximation
algorithms.

A necessary and sufficient condition for a complete graph with $n$ vertices to
have an $L$-cycle cover is that there exist (not necessarily distinct) lengths
$\lambda_1, \ldots, \lambda_k \in L$ for some $k \in \nat$ with
$\sum_{i=1}^k \lambda_i = n$. We call such an $n$ \bemph{$L$-admissible} and
define $\close L = \{n \mid \text{$n$ is $L$-admissible}\}$. Although $L$ can be
arbitrarily complicated, $\close L$ always allows efficient membership testing.

\begin{lemma}
\label{gcc:lem:replacement}
  For all $L \subseteq \nat$, there exists a finite set $L' \subseteq L$ with
  $\close{L'} = \close L$.
\end{lemma}

\begin{proof}
  Let $L_{\leq \ell} = \{n \in L\mid n \leq \ell\} \subseteq L$. Let
  $g_L \in \nat$ be the greatest common divisor of all numbers in $L$. There
  exists an $\ell_0 \in L$ such that $g_L$ is also the greatest common divisor
  of $L_{\leq \ell_0}$.

  If $g_L \in L$, then $\close{\{g_L\}} = \close L$, and we are done. Thus, we
  assume $g_L \notin L$. There exist $\xi_1, \ldots, \xi_k \in \integer$ and
  $\lambda_1, \ldots, \lambda_k \in L_{\leq \ell_0}$ for some $k \in \nat$ with
  $\sum_{i=1}^k \xi_i \lambda_i = g_L$. Let
  $\xi = \min_{1 \leq i \leq k} \xi_i$. We have $\xi < 0$ since $g_L \notin L$.
  Choose any $\lambda \in L_{\leq \ell_0}$ and let $\ell = -\xi\lambda \cdot
  \sum_{i=1}^k \lambda_i$. Let $n \in \close L$ with $n \geq \ell$, let
  $m = \bmod(n-\ell, \lambda)$, and let $s = \left\lfloor \frac{n-\ell}\lambda
  \right\rfloor$. We can write $n$ as
  \[
  n = \lambda s + m + \ell
    = \lambda s + \frac{m}{g_L} \cdot \sum_{i=1}^k \xi_i \lambda_i
                - \lambda \xi \cdot \sum_{i=1}^k \lambda_i 
    = \lambda s + \sum_{i=1}^k (m \xi_i - \lambda \xi) \cdot \lambda_i .
  \]
  Since $m < \lambda$ and $\xi_i \geq \xi< 0$, we have
  $(m \xi_i - \lambda \xi) \geq 0$ for all $i$. Hence, $\close{L_{\leq \ell_0}}$
  contains all elements $n \in \close L$ with $n \geq \ell$. Elements of
  $\close L$ smaller than $\ell$ are contained in
  $\close{L_{\leq \ell}} \supseteq \close{L_{\leq \ell_0}}$. Hence,
  $\close{L_{\leq \ell}} = \close L$ and $L' = L_{\leq \ell}$ is the finite set
  we are looking for.
\end{proof}

For every fixed $L$, we can not only test in time polynomial in $n$ whether $n$
is $L$-admissible, but we can, provided that $n \in \close L$, also find numbers
$\lambda_1, \ldots, \lambda_k \in L'$ that add up to $n$, where $L' \subseteq L$
denotes a finite set with $\close L = \close{L'}$. This can be done via dynamic
programming in time $O(n \cdot |L'|)$, which is $O(n)$ for fixed $L$.

Although $\close L = \close{L'}$, there are clearly graphs for which the weights
of an optimal $L$-cycle cover and an optimal $L'$-cycle cover differ: Let
$\lambda \in L \setminus L'$ and consider a $\lambda$-vertex graph where all
edge weights are zero except for one Hamiltonian cycle of weight one edges.
However, this does not matter for our approximation algorithms.

The two approximation algorithms presented in Sections~\ref{ssec:undirectedalg}
and~\ref{ssec:directedalg} are based on a decomposition technique for cycle
covers presented in Section~\ref{ssec:decomposition}.

%%%%%%%%%%%%%%%%%%%%%%%%%%%%%%%%%%%%%
\subsection{Decomposing Cycle Covers}
\label{ssec:decomposition}

In this section, we present a decomposition technique for cycle covers. The
technique can be applied to cycle covers of undirected graphs but also to
directed cycle covers that do not contain 2-cycles.

A \bemph{single} is a single edge (or a path of length one) in a graph, while a
\bemph{double} is a path of length two. Our aim is to decompose a cycle cover
$C$ on $n$ vertices into roughly $n/6$ singles, $n/6$ doubles, and $n/6$
isolated vertices. If $n$ is not divisible by six, we replace $n/6$ by
$\lfloor n/6 \rfloor$ or $\lceil n/6 \rceil$: If $n=6k + \ell$ for
$k, \ell \in \nat$ and $\ell \leq 5$, then we take $k + \alpha_\ell$ singles and
$k+ \beta_\ell$ doubles, where $\alpha_\ell$ and $\beta_\ell$ are given in
Table~\ref{tab:alphabeta}. Thus, we retain half of the edges of $C$. We aim to
decompose the cycle covers such that at least half of the weight of the cycle
cover is preserved.

\begin{table}[t] \small
\centering
\begin{tabular}{|l||c|c|c|c|c|c|}        \hline
$\ell$        & 0 & 1 & 2 & 3 & 4 & 5 \\ \hline \hline
$\alpha_\ell$ & 0 & 1 & 1 & 0 & 0 & 1 \\ \hline 
$\beta_\ell$  & 0 & 0 & 0 & 1 & 1 & 1 \\ \hline 
\end{tabular}
\caption{A cycle cover on $n = 6k+\ell$ vertices will be decomposed into
$k + \alpha_\ell$ singles and $k + \beta_\ell$ doubles.}
\label{tab:alphabeta}
\end{table}

The reason why we decompose cycle covers into singles and doubles is the
following: We cannot decompose them into longer paths in general since this does
not work for $\{3\}$-cycle covers. If we restricted ourselves to decomposing the
cycle covers into singles only, then $3$-cycles would limit the weight
preserved: We would retain only one third of the edges of the 3-cycles, thus at
most one third of their weight in general. Finally, if we restricted ourselves
to doubles, then $5$-cycles would limit the weight we could obtain since we
would retain only two of their five edges.

In our approximation algorithms, we exploit the following observation: If every
cycle cover on $n$ vertices can be decomposed into $\alpha$ singles and $\beta$
doubles, then, for every $L$, every $L$-cycle cover on $n$ vertices can be
decomposed in the same way. This implies that we can build cycle covers from
such a decomposition: Given $\alpha$ singles and $\beta$ doubles, and
$n-2 \alpha - 3 \beta$ isolated vertices, we can join them to form an $L$-cycle
cover. (The only restriction is that $n$ must be $L$-admissible.)

Let us now state the decomposition lemma.

\begin{lemma}
\label{lem:decomposition}
  Let $C=(V, E)$ be a cycle cover on $n = 6 k + \ell$ vertices such that the
  length of each cycle is at least three. Let $w : E \rightarrow \nat$ be an
  edge weight function.

  Then there exists a decomposition $D \subseteq E$ of $C$ such that $(V, D)$
  consists of vertex-disjoint $k + \alpha_\ell$ singles, $k + \beta_\ell$
  doubles, and $n - 5k - 3 \beta_\ell - 2 \alpha_\ell$ isolated vertices and
  $w(D) \geq w(E)/2$, where $\alpha_\ell$ and $\beta_\ell$ are given in
  Table~\ref{tab:alphabeta}.

  The decomposition can be done in polynomial time.
\end{lemma}

\begin{figure}[t]
\centering
\subfigure[A cycle cover.]{%
~ \qquad
\scalebox{0.9}{\includegraphics{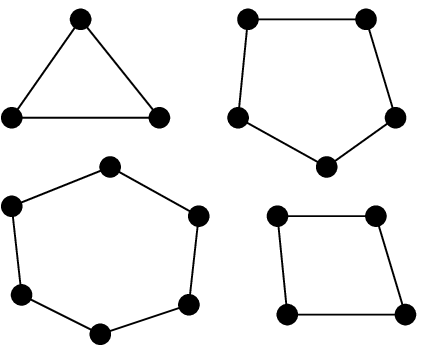}}
\qquad ~}
\qquad
\subfigure[A decomposition of the cycle cover.]{%
~ \qquad
\scalebox{0.9}{\includegraphics{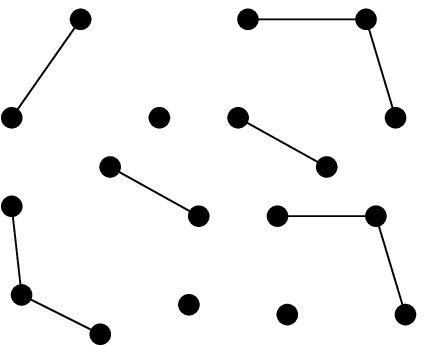}}
\qquad ~}
\caption{An example of a decomposition according to
Lemma~\ref{lem:decomposition}.}
\label{fig:decomposition}
\end{figure}

Figure~\ref{fig:decomposition} illustrates how a cycle cover is decomposed into
singles and doubles.

Let us first prove some helpful lemmas.

\begin{lemma}
\label{cla:basicdecomp}
  Let $\lambda, \alpha, \beta \in \nat$ with $\alpha + 2 \beta \geq \lambda/2$
  and $2 \alpha + 3 \beta \leq \lambda$. Then every cycle $c$ of length
  $\lambda$ can be decomposed into $\alpha$ singles and $\beta$ doubles such
  that the weight of the decomposition is at least $w(c)/2$.
\end{lemma}

\begin{proof}
  Every single involves two vertices of $c$ while every double involves three
  vertices. Thus, $2 \alpha + 3 \beta \leq \lambda$ is a necessary condition for
  $c$ being decomposable into $\alpha$ singles and $\beta$ doubles. It is also a
  sufficient condition.

  We assign an arbitrary orientation to $c$. Let $e_0, \ldots, e_{\lambda-1}$ be
  the consecutive edges of $c$, where $e_0$ is chosen uniformly at random among
  the edges of $c$. We take $\alpha$ singles
  $e_{0}, e_{2}, \ldots, e_{2\alpha -2}$ and $\beta$ doubles $(e_{2 \alpha},
  e_{2 \alpha +1}), (e_{2 \alpha+3}, e_{2 \alpha +4}), \ldots,
  (e_{2 \alpha + 3 \beta -3}, e_{2 \alpha + 3 \beta -2})$. Since
  $2 \alpha + 3 \beta \leq \lambda$, this is a feasible decomposition. The
  probability that any fixed edge of $c$ is included in the decomposition is
  $\frac{\alpha + 2\beta}{\lambda}$. Thus, the expected weight of the
  decomposition is $\frac{\alpha + 2\beta}{\lambda} \cdot w(c) \geq w(c)/2$.
\end{proof}

\begin{lemma}
\label{cla:plussix}
  Let $\lambda \in \nat$. Suppose that every cycle $c$ of length $\lambda$ can
  be decomposed into $\alpha$ singles and $\beta$ doubles of weight at least
  $w(c)/2$. Then every cycle $c'$ of length $\lambda+6$ can be decomposed into
  $\alpha+1$ singles and $\beta+1$ doubles of weight at least $w(c')/2$.
\end{lemma}

\begin{proof}
  We have $\alpha + 2 \beta \geq \lambda /2$ and
  $2 \alpha + 3 \beta \leq \lambda$. Thus,
  $\alpha +1 + 2 (\beta +1) \geq (\lambda +6)/2$ and
  $2 (\alpha +1) + 3 (\beta+1) \leq \lambda +6$.
  The lemma follows from Lemma~\ref{cla:basicdecomp}.
\end{proof}

Lemma~\ref{cla:plussix} also holds if we consider more than one cycle: Assume
that every collection of $k$ cycles of lengths $\lambda_1, \ldots, \lambda_k$
can be decomposed into $\alpha$ singles and $\beta$ doubles such that the weight
of the decomposition is at least half the weight of the cycles. Then $k$ cycles
of lengths $\lambda_1+6, \lambda_2, \ldots, \lambda_k$ can be decomposed into
$\alpha+1$ singles and $\beta +1$ doubles such that also at least half of the
weight of the cycles is preserved. Due to Lemma~\ref{cla:plussix}, we can
restrict ourselves to cycles of length at most eight in the following. The
reason for this is the following: If we know how to decompose cycles of length
$\lambda$, then we also know how to decompose cycles of length
$\lambda +6, \lambda + 12, \ldots$ from Lemma~\ref{cla:plussix}.

We are now prepared to prove Lemma~\ref{lem:decomposition}.

\begin{proof}[Proof of Lemma~\ref{lem:decomposition}]
  We prove the lemma by induction on the number of cycles. As the induction
  basis, we consider a cycle cover consisting of either a single cycle or of two
  odd cycles. Due to Lemma~\ref{cla:plussix}, we can restrict ourselves to
  considering cycles of length at most eight. Tables~\ref{subfig:baseonecycle}
  and~\ref{subfig:basetwocycles} show how to decompose a single cycle and two
  odd cycles, respectively. We always perform the decomposition such that the
  weight preserved is maximized. In particular, if there are two odd cycles of
  different length, we have two options in how to decompose these cycles, and we
  choose the one that yields the larger weight. Overall, we obtain a
  decomposition with an appropriate number of singles and doubles that preserves
  at least one half of the weight.

\newlength{\tolengthtable}
\begin{table}
\small \settowidth{\tolengthtable}{to}
\centering
\subtable[One cycle.]{%
\label{subfig:baseonecycle}%
\begin{tabular}{|c|c|c|c|} \hline
\textbf{length} & {\boldmath $\ell$} &
{\boldmath $\alpha$}& {\boldmath $\beta$} \\ \hline
3 & 3 & 0 & 1 \\ \hline
4 & 4 & 0 & 1 \\ \hline
5 & 5 & 1 & 1 \\ \hline
6 & 0 & 1 & 1 \\ \hline
7 & 1 & 2 & 1 \\ \hline
8 & 2 & 2 & 1 \\ \hline
\end{tabular}
}
\quad
\subtable[Two odd cycles.]{%
\label{subfig:basetwocycles}%
\begin{tabular}{|c|c|c|c|l|} \hline
\textbf{lengths} & {\boldmath $\ell$} &
{\boldmath $\alpha$}& {\boldmath $\beta$} &
\textbf{decomposition} \\ \hline
3 + 3 & 0 & 1 & 1 &
\hspace{\tolengthtable} 3$\leadsto$(1,0) + 3$\leadsto$(0,1) \\ \hline
3 + 5 & 2 & 2 & 1 &
\hspace{\tolengthtable} 3$\leadsto$(1,0) + 5$\leadsto$(1,1) \\ 
& &  &  & or 3$\leadsto$(0,1) + 5$\leadsto$(2,0) \\ \hline
3 + 7 & 4 & 1 & 2 &
\hspace{\tolengthtable}  3$\leadsto$(1,0) + 7$\leadsto$(0,2) \\ 
& &  &  & or 3$\leadsto$(0,1) + 7$\leadsto$(1,1) \\ \hline
5 + 5 & 4 & 1 & 2 &
\hspace{\tolengthtable} 5$\leadsto$(0,1) + 5$\leadsto$(1,1) \\ \hline
5 + 7 & 0 & 2 & 2 &
\hspace{\tolengthtable} 5$\leadsto$(2,0) + 7$\leadsto$(0,2) \\ 
& &  &  & or 5$\leadsto$(1,1) + 7$\leadsto$(1,1) \\ \hline
7 + 7 & 2 & 3 & 2 &
\hspace{\tolengthtable}  7$\leadsto$(1,1) + 7$\leadsto$(2,1) \\  \hline
\end{tabular}}
\caption{The induction basis. The columns $\alpha$ and $\beta$ show the number
of singles and doubles needed, respectively. We denote by
$\lambda \leadsto (\alpha,\beta)$ that a $\lambda$-cycle is decomposed into
$\alpha$ singles and $\beta$ doubles. If there are two lines for a case, then
the option that yields more weight is chosen.}
\label{fig:basistable}
\end{table}

  As the induction hypothesis, we assume that the lemma holds if the number of
  cycles is less than $r$. Assume that we have a cycle cover $C$ consisting of
  $r$ cycles. Let $n = 6 k + \ell$ for the number of its vertices for
  $k, \ell \in \nat$ and $\ell \leq 5$. We remove either an even cycle or two
  odd cycles. In the following, let $C'$ be the new cycle cover obtained by
  removing one or two cycles from $C$. A little more care is needed than in the
  induction basis: Consider for instance the case of removing a $4$-cycle. If
  $\ell = 4$, then $C$ has to be decomposed into $k$ singles and $k+1$ doubles,
  while we have to take $k$ singles and $k$ doubles from $C'$. Thus, the 4-cycle
  has to be decomposed into a double. But if $\ell = 1$, then we need $k+1$
  singles and $k$ doubles from $C$ and $k-1$ singles and $k$ doubles from $C'$.
  Thus, the $4$-cycle has to be decomposed into two singles. Overall, the
  $4$-cycle has to be decomposed into a double if $\ell \in \{0,3,4,5\}$ and
  into two singles if $\ell \in \{1,2\}$. Similar case distinctions hold for all
  other cases. How to remove one even or two odd cycles is shown in
  Tables~\ref{subfig:onecycle} and~\ref{subfig:twocycles}, respectively.

\begin{table}
\small \centering
\subtable[Removing an even cycle.]{%
\label{subfig:onecycle}%
\begin{tabular}{|c|l|c|c|} \hline
\textbf{length} & {\boldmath $\ell$} &
{\boldmath $\alpha$}& {\boldmath $\beta$}  \\ \hline
4 & 0,3,4,5 & 0 & 1 \\ \hline
4 & 1,2 & 2  & 0 \\ \hline
6 & all & 1 & 1 \\ \hline
8 & 0,1,2,5 & 2 & 1 \\ \hline
8 & 3,4 & 0 & 2 \\ \hline
\end{tabular}
}
\quad
\subtable[Removing two odd cycles.]{%
\label{subfig:twocycles}%
\begin{tabular}{|c|l|c|c|l|} \hline
\textbf{lengths} & {\boldmath $\ell$} &
{\boldmath $\alpha$}& {\boldmath $\beta$} &
\textbf{decomposition} \\ \hline
3 + 3 & all & 1 & 1 &
\hspace{\tolengthtable} 3$\leadsto$(1,0) + 3$\leadsto$(0,1) \\ \hline
3 + 7 & 0,3,4,5 & 1 & 2 &
\hspace{\tolengthtable} 3$\leadsto$(1,0) + 7$\leadsto$(0,2) \\ 
& &  &  &or  3$\leadsto$(0,1) + 7$\leadsto$(1,1) \\ \hline
3 + 7 & 1,2 & 3 & 1 &
\hspace{\tolengthtable} 3$\leadsto$(1,0) + 7$\leadsto$(2,1) \\ 
& &  &  &or  3$\leadsto$(0,1) + 7$\leadsto$(3,0) \\ \hline
5 + 5 & 0,3,4,5 & 1 & 2 &
\hspace{\tolengthtable} 5$\leadsto$(0,1) + 5$\leadsto$(1,1) \\ \hline
5 + 5 & 1,2 & 3 & 1 &
\hspace{\tolengthtable} 5$\leadsto$(2,0) + 5$\leadsto$(1,1) \\ \hline
5 + 7 & all & 2 & 2 &
\hspace{\tolengthtable} 5$\leadsto$(2,0) + 7$\leadsto$(0,2) \\ 
& &  &  & or 5$\leadsto$(1,1) + 7$\leadsto$(1,1) \\ \hline
7 + 7 & 0,1,2,5 & 3 & 2 &
\hspace{\tolengthtable}  7$\leadsto$(1,1) + 7$\leadsto$(2,1) \\  \hline
7 + 7 & 3,4 & 1 & 3 &
\hspace{\tolengthtable}  7$\leadsto$(1,1) + 7$\leadsto$(0,2) \\  \hline
\end{tabular}}
\caption{Induction step.}
\end{table}

  To complete the proof, we have to deal with the case of a $3$- and a
  $5$-cycle, which is slightly more complicated and not covered by
  Table~\ref{subfig:twocycles}. We run into trouble if, for instance, $\ell =3$.
  In this case, we have to take two doubles. If the 5-cycle is much heavier than
  the 3-cycle, then it is impossible to preserve half of the weight of the two
  cycles. But we can avoid this problem: As long as there is an even cycle, we
  decompose this one. After that, as long as there are at least three odd
  cycles, we can choose two of them such that we do not have a pair of one
  $(3+6\xi)$-cycle and one $(5+6\xi')$-cycle for some $\xi, \xi' \in \nat$. The
  only situation in which it can happen that we cannot avoid decomposing a
  $(3+6\xi)$-cycle and a $(5+6\xi')$-cycle is when there are only two cycles
  left. In this case, we have $\ell = 2$, and we have treated this case already
  in the induction basis.
\end{proof}

If we consider directed graphs where $2$-cycles can also occur, only one third
of the weight can be preserved. This can be done by decomposing the cycle cover
into a matching of cardinality $\lceil n/3\rceil$. (Every $\lambda$-cycle for
can be decomposed into a matching of size up to
$\lfloor \lambda/2 \rfloor \geq \lceil \lambda/3\rceil$. The bottleneck are
$3$-cycles, which yield only one edge.)

An obvious question is whether the decomposition lemma can be improved in order
to preserve more than half of the weight or more than one third of the weight if
we additionally allow $2$-cycles. Unfortunately, this is not the case.

A generic decomposition lemma states the following: For every $n \in \nat$,
every $k$-cycle cover (for $k \in \{2,3\}$) on $n$ vertices can be decomposed
into $\alpha$ singles and $\beta$ doubles such that at least a fraction $r$ of
the weight of the cycle cover is preserved. (As already mentioned, longer paths
are impossible due to $3$-cycles.) Lemma~\ref{lem:decomposition} instantiates
this generic lemma with $\alpha \approx n/6$, $\beta \approx n/6$, and
$r = 1/2$. In case of the presence of $2$-cycles, we have sketched a
decomposition with $\alpha \approx n/3$, $\beta = 0$, and $r = 1/3$.

\begin{lemma}
\label{lem:tight}
  No decomposition technique for $3$-cycle covers can in general preserve more
  than one half of the weight of the $3$-cycle covers.

  Furthermore, no decomposition technique for $2$-cycle covers can in general
  preserve more than one third of the weight of the $2$-cycle covers.
\end{lemma}

\begin{proof}
  We exploit the fact that the fraction of edges that are preserved in a cycle
  cover decomposition is a lower bound for the fraction of the weight that can
  be preserved.

  Since, in particular, $\{3\}$-cycle covers have to be decomposed, we cannot
  decompose the cycle cover into paths of length more than two. Now consider
  decomposing a $\{4\}$-cycle cover. Since paths of length $3$ are not allowed,
  we have to discard two edges of every $4$-cycle. Thus, at most $2$ edges of
  every $4$-cycle are preserved, which proves the first part of the lemma.

  The second part follows analogously by considering $3$-cycles and observing
  that paths of length two or more are not allowed.
\end{proof}

Overall, Lemma~\ref{lem:tight} shows that every approximation algorithm for
\maxug L or \maxdg L that works for arbitrary sets $L$ and is purely
decomposition-based achieves approximation ratios of at best $2$ or $3$,
respectively. We achieve an approximation ratio of $8/3 < 3$ for \maxdg L by
paying special attention to $2$-cycles (Section~\ref{ssec:directedalg}).

%%%%%%%%%%%%%%%%%%%%%%%%%%%%%%%%%%%%
\subsection{Undirected Cycle Covers}
\label{ssec:undirectedalg}

Our approximation algorithm for \maxug L (Algorithm~\ref{algo:undirected})
directly exploits Lemma~\ref{lem:decomposition}.

\begin{theorem}
  Algorithm~\ref{algo:undirected} is a factor $2$ approximation algorithm for
  \maxug L for all $L \subseteq \unduniv$. Its running-time is $O(n^3)$.
\end{theorem}

\begin{proof}
  If $L$ is infinite, we replace $L$ by a finite set $L' \subseteq L$ with
  $\close{L'} = \close L$ according to Lemma~\ref{gcc:lem:replacement}.
  Algorithm~\ref{algo:undirected} returns $\bot$ if and only if
  $n \notin \close L$. Otherwise, an $L$-cycle cover $C^\apx$ is returned.
  Let $C^\star$ denote an $L$-cycle cover of maximum weight of $G$. We have
  $w(C^\star) \leq w(C^\init) \leq 2 \cdot w(D) \leq 2 \cdot w(C^\apx)$. The
  first inequality holds because $L$-cycle covers are special cases of cycle
  covers. The second inequality holds due to the decomposition lemma
  (Lemma~\ref{lem:decomposition}). The last inequality holds since no weight is
  lost during the joining. Overall, the algorithm achieves an approximation
  ratio of $2$.

  The running-time of the algorithm is dominated by the time needed to compute
  the initial cycle cover, which is
  $O(n^3)$~\cite[Chapter~12]{AhujaEA:NetworkFlows:1993}.
\end{proof}

\begin{algorithm}[t]
\begin{algorithmic}[1]
\Input undirected complete graph $G = (V, E)$, $|V| = n$;
       edge weights $w: E \rightarrow \nat$
\Output an $L$-cycle cover $C^\apx$ of $G$ if $n$ is $L$-admissible,
        $\bot$ otherwise
\If{$n \notin \close L$}
  \State \textbf{return} $\bot$
\EndIf
\State compute a cycle cover $C^\init$ in $G$ of maximum weight
\State decompose $C^\init$ into a set $D \subseteq C^\init$ of edges according
       to Lemma~\ref{lem:decomposition}
\State join the singles and doubles in $D$ to obtain an $L$-cycle cover $C^\apx$
\State \textbf{return} $C^\apx$
\end{algorithmic}
\caption{A $2$-approximation algorithm for \maxug L.}
\label{algo:undirected}
\end{algorithm}

%%%%%%%%%%%%%%%%%%%%%%%%%%%%%%%%%%
\subsection{Directed Cycle Covers}
\label{ssec:directedalg}

In the following, let $C^\opto$ be an $L$-cycle cover of maximum weight. Let
$w_\lambda$ denote the weight of the $\lambda$-cycles in $C^\opto$, \ie,
$w(C^\opto) = \sum_{\lambda \geq 2} w_{\lambda}$.

\begin{algorithm}[t]
\begin{algorithmic}[1]
\Input directed complete graph $G = (V, E)$, $|V| = n$;
       edge weights $w: E \rightarrow \nat$
\Output an $L$-cycle cover $C^\apx$ of $G$ if $n$ is $L$-admissible,
        $\bot$ otherwise
\If{$n \notin \close L$}
  \State \textbf{return} $\bot$
\EndIf
\If{$2 \in L$ and $3 \in L$}
  \State compute a cycle cover $C^\init$ (without restrictions)
  \ForAll{even cycles $c$ of $C^\init$}
    \State take every other edge of $c$ such that at least one half of $c$'s
           weight is preserved
    \State add the converse edges to obtain $2$-cycles; add these cycles to
           $C^\apx$
  \EndFor
  \ForAll{odd cycles $c$ of $C^\init$}
    \State take every other edge and one path of length two of $c$ such that at
           least one half of $c$'s weight is preserved
    \State add edges to obtain $2$-cycles plus one $3$-cycle; add these cycles
           to $C^\apx$
  \EndFor
\ElsIf{$2 \in L$, $3 \notin L$}
  \State compute a matching $M$ of maximum weight of cardinality at most
         $D(n,L)$
  \State join the edges of $M$ to form an $L$-cycle cover $C^\apx$
\Else ~($2 \notin L$)
  \State compute a $4/3$-approximation $C^\init_3$ to an optimal $3$-cycle cover
  \State decompose $C^\init_3$ into a set $D \subseteq C^\init_3$ of edges
         according to Lemma~\ref{lem:decomposition}
  \State join the singles and doubles in $D$ to obtain an $L$-cycle $C^\apx$
\EndIf
\State \textbf{return} $C^\apx$
\end{algorithmic}
\caption{A factor $8/3$ approximation algorithm for \maxdg L.}
\label{algo:directed}
\end{algorithm}

We distinguish three cases: First, $2 \notin L$, second, $2 \in L$ and
$3 \notin L$, and third, $2,3 \in L$.

We use the decomposition lemma (Lemma~\ref{lem:decomposition}) only if
$2 \notin L$. In this case, the weight of an optimal $L$-cycle cover is at most
the weight of an optimal $3$-cycle cover $C^\opto_3$. Thus, we proceed as
follows: First, we compute a $4/3$ approximation $C^\init_3$ for \maxdg 3, which
can be done by using the algorithm of Bl\"aser et
al.~\cite{BlaeserEA:MetricMaxATSP:2005}. We have
$w(C^\init_3) \geq \frac 34 \cdot w(C^\opto_3) \geq \frac 34 \cdot w(C^\opto)$.
Now we decompose $C^\init_3$ into a collection $D$ of singles and doubles
according to Lemma~\ref{lem:decomposition}. Finally, we join the singles,
doubles, and isolated vertices of $D$ to form an $L$-cycle cover $C^\apx$. We
obtain a factor $8/3$ approximation for the case that $2 \notin L$:
\[
w(C^\apx) \geq w(D)
          \geq \frac 12 \cdot w(C^\init_3)
          \geq \frac 38 \cdot w(C^\opto).
\]

Now we consider the case that $2 \in L$ and $3 \notin L$. In this case, a
matching-based algorithm achieves an approximation ratio of $5/2$: We compute a
matching of a certain cardinality, which we will specify in a moment, and then
we join the edges of the matching to obtain an $L$-cycle cover. The cardinality
of the matching is chosen such that an $L$-cycle cover can be built from such a
matching. A $\lambda$-cycle yields a matching of cardinality
$\lfloor \lambda/2 \rfloor$. Thus, a matching of cardinality $d$ in a graph of
$n$ vertices can be extended to form an $L$-cycle cover if and only if
$d \leq D(n, L)$, where 
\[
D(n, L) =  \max\left\{\sum_{i=1}^k \lfloor\lambda_i/2\rfloor
                      \mid k \in \nat,
                           \sum_{i=1}^k \lambda_i = n \text{, and }
                           \lambda_i \in L\text{ for $1 \leq i \leq k$}\right\}
\leq \frac n2 .
\]
Given $L$, we can compute $D(n,L)$ using dynamic programming. Let us now
estimate the weight of a matching of cardinality at most $D(n,L)$ that has
maximum weight among all such matchings. From $C^\opto$, we obtain a matching
with a weight of at least
\[
\sum_{\lambda\geq 2} \frac{1}{\lambda}
   \cdot \left\lfloor \frac{\lambda}2 \right\rfloor
   \cdot w_\lambda
\geq \sum_{\lambda \geq 2} \frac 25 \cdot w_\lambda
 =   \frac 25 \cdot w(C^\opto) .
\]
The reason is that $w_3 = 0$ because $3 \notin L$ and that
$\min_{\lambda \in \{2,4,5,6,7,\ldots\}} \frac{1}{\lambda} \cdot
\lfloor \lambda/2 \rfloor \geq 2/5$. Thus, by computing a maximum-weight
matching $M$ of cardinality at most $D(n,L) \geq 2n/5$ and joining the edges to
form an $L$-cycle cover $C^\apx$, we obtain a factor $5/2$ approximation.

\begin{figure}[t]
\centering
\subfigure[Initial cycle cover $C^\init$.]{%
~ ~
\scalebox{0.9}{\includegraphics{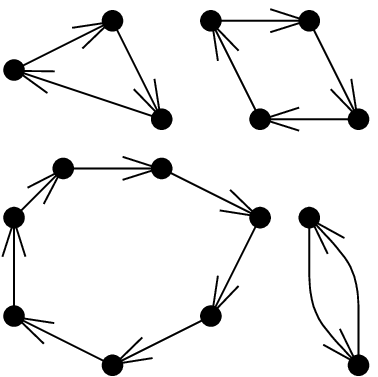}}
 ~ ~}
\quad
\subfigure[Decomposition of $C^\init$.]{%
~ ~
\scalebox{0.9}{\includegraphics{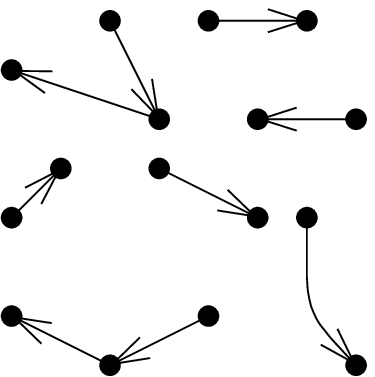}}
 ~ ~}
\quad
\subfigure[$\{2,3\}$-cycle cover $C^\apx$.]{%
~ ~
\scalebox{0.9}{\includegraphics{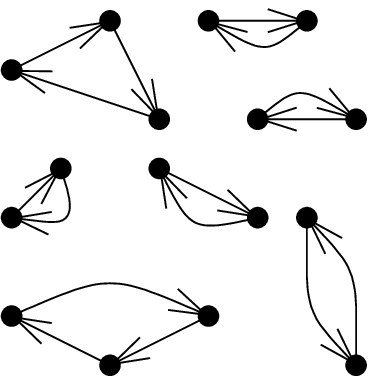}}
 ~ ~}
\caption{Sketch of the algorithm for $\{2,3\} \subseteq L$.}
\label{fig:directeddecomp}
\end{figure}

What remains to be considered is the case that $2,3  \in L$. In this case, we
start by computing an initial cycle cover $C^\init$ (without any restrictions).
Then we do the following: For every even cycle, we take every other edge such
that at least one half of its weight is preserved. For every edge thus obtained,
we add the converse edge to obtain a collection of $2$-cycles. For every odd
cycle, we take every other edge and one path of length two such that at least
half of the weight is preserved. Then we add edges to obtain $2$-cycles and one
$3$-cycle. In this way, we obtain a $\{2,3\}$-cycle cover $C^\apx$, which is
also an $L$-cycle cover. We have
$w(C^\apx) \geq \frac 12 \cdot w(C^\init) \geq \frac 12 \cdot w(C^\opto)$. Figure~\ref{fig:directeddecomp} shows an example.

Our approximation algorithm is summarized as Algorithm~\ref{algo:directed}. The
running-time of the algorithm of Bl\"aser et al.\ is
polynomial~\cite{BlaeserEA:MetricMaxATSP:2005} and all other steps can be
executed in polynomial time as well. Thus, the running-time of
Algorithm~\ref{algo:directed} is also polynomial.

\begin{theorem}
  Algorithm~\ref{algo:directed} is a factor $8/3$ approximation algorithm for
  \maxug L for all non-empty sets $L \subseteq \diruniv$. Its running-time is
  polynomial.
\end{theorem}

%%%%%%%%%%%%%%%%%%%%%
\section{Conclusions}
\label{sec:concl}
%%%%%%%%%%%%%%%%%%%%%

\begin{table}[t]
\small \centering 
\subtable[Undirected cycle covers.]{%
\begin{tabular}{|l||l|l|l|}
  \hline
      & \bemph{\ucc L} & \bemph{\maxu L} & \bemph{\maxug L} \\ \hline \hline
  \bemph{$\oL = \emptyset$}
      & in \DP         & in \PO          & in \PO           \\ \hline
  \bemph{$\oL = \{3\}$}
      & in \DP         & in \PO          &                  \\ \hline
  \bemph{$\oL = \{4\}, \{3,4\}$}
      &                &                 & \APX-complete    \\ \hline
  \bemph{$\oL \not\subseteq \{3,4\}$}
      & \NP-hard       & \APX-hard       & \APX-hard        \\ \hline
\end{tabular}}
\\
\subtable[Directed cycle covers.]{%
\begin{tabular}{|l||l|l|l|}
  \hline
      & \bemph{\dcc L} & \bemph{\maxd L} & \bemph{\maxdg L} \\ \hline \hline
  \bemph{$L = \{2\}, \diruniv$}
      & in \DP         & in \PO          & in \PO           \\ \hline
  \bemph{$L \notin \{\{2\}, \diruniv\}$}
      & \NP-hard       & \APX-hard       & \APX-hard        \\ \hline
\end{tabular}}
\caption{The complexity of computing $L$-cycle covers.}
\label{tab:newworld}
\end{table}

For almost all $L$, finding $L$-cycle covers is \NP-hard and finding $L$-cycle
covers of maximum weight is \APX-hard. Table~\ref{tab:newworld} shows an
overview. Although this shows that computing restricted cycle covers is
generally very hard, we have proved that $L$-cycle covers of maximum weight can
be approximated within a constant factor in polynomial time for all~$L$.

For directed graphs, we have settled the complexity: If $L = \{2\}$ or
$L = \diruniv$, then \dcc L, \maxd L, and \maxdg L are solvable in polynomial
time, otherwise they are intractable. For undirected graphs, the status of only
five cycle cover problems remains open: \ucc L and \maxu L for
$\oL = \{4\},  \{3,4\}$ and \maxug 4.

There are some reasons for optimism that \ucc L and \maxu L for
$\oL = \{4\},  \{3,4\}$ are solvable in polynomial time:
Hartvigsen~\cite{Hartvigsen:SquareFree:2006} devised a polynomial-time algorithm
for finding $\overline{\{4\}}$-cycle covers in bipartite graphs (forbidding
$3$-cycles does not change the problem for bipartite graphs). Moreover, there
are augmenting path theorems for $L$-cycle covers for all $L$ with
$\oL \subseteq \{3,4\}$~\cite{Russell:Master:2001}, which includes the two cases
that are known to be polynomial-time solvable. Augmenting path theorems are
often a building block for matching algorithms. But there are also augmenting
path theorems for $L \subseteq \{3,4\}$~\cite{Russell:Master:2001}, even though
these $L$-cycle cover problems are intractable.

%%%%%%%%%%%%%%%%%%%%%%%%%%
\section*{Acknowledgments}
%%%%%%%%%%%%%%%%%%%%%%%%%%

I thank Jan Arpe and Martin B\"ohme for valuable discussions and comments.

%%%%%%%%%%%%%%
% Bibliography
%%%%%%%%%%%%%%

\end{document}